# Extended Einstein diffusion-mobility equation for two-dimensional Schrödinger-type quantum materials

K. Navamani*
*Theoretical Sciences Unit*
*Jawaharlal Nehru Centre for Advanced Scientific Research*
*Jakkur-560064, Bangalore, INDIA*
*e-mail address: pranavam5q@gmail.com*

Dedicated to Professor Swapan K. Pati

We present the exact analytical equation of diffusion-mobility for two-dimensional (2D) Schrödinger type transport systems, from molecules to materials. The density of electronic states in such Schrödinger systems pertains to the 2D non-relativistic carrier dynamics. We implement the Gaussian function into carrier density derivation; accordingly we develop the electronic compressibility and diffusion-mobility for both the generic and the degenerate Fermi systems. This model is originally developed from generalized Einstein relation, along with concern about the thermodynamic effects on many-body interactions. The effect of interactions is included through the imperfect Fermi-gas entropy function. Our extended model explains the cooperative behavior of thermal and electronic counterparts on diffusion-mobility in disordered systems at wide temperature range. Using earlier experimental and theoretical results, we have shown the validity of our extended Einstein model for different 2D degenerate systems. The results validate the original Einstein equation at certain sets of temperature and chemical potential values for different Gaussian variances. Beyond those combinations, the deviation is observed. At very low temperature, the diffusion-mobility depends only on chemical potential, which is the extended Einstein equation for ideal quantum materials.

## I. INTRODUCTION

The renowned Einstein's diffusion-mobility equation is widely used to investigate the semiconducting properties of highly disordered systems like, molecular solids [1, 2]. Various studies in the past five decades emphasize two important corollaries; (i) the Einstein equation works pretty well only in nondegenerate classical systems at equilibrium in high temperature domain of $T > 150K$ [3, 4], (ii) on the other hand, it is not valid for the high charge density (degenerate) limit in quantum materials, even at equilibrium condition [5-10]. Moreover, our recent analysis on carrier drift energy-current density in the organic molecules manifests the deviation of Einstein equation under the applied electric field situations [11, 12]. In this study, the obtained ideality factor through Shockley diode equation for different molecules is in the range of 1.8-2.0, which is in agreement with the earlier reports [13, 14]. Besides the organic solids, the estimated ideality factor for the periodic systems shows high values and it apparently varies with the chemical potential (Fermi energy) by carrier doping, in which Einstein equation fails miserably [8, 9, 15-17]. In principle, the diffusion-mobility equation is associated with the carrier density and the electronic density of states (or charge compressibility) of the materials. In this case, compressibility ensures the measurement of many-body interactions [18-20]. In the extreme quantum degenerate regimes under the strong electric or magnetic field conditions, the shifting of chemical potential suggests the formation of many-body incompressible states, which is direct evidence of density flux [21-23]. Here, the inhomogeneity in carrier puddles facilitates the quantum diffusion. Basically, the classical Einstein relation is directly related to thermal energy ($k_B T$) which shows the linear dependency on temperature, but there is no electronic counterpart like, DOS or compressibility for degenerate nanosystems to drag the quantum phenomena on this equation. Importantly, the missing of many-body DOS information for highly degenerate cases indicates the necessity of revisiting the Einstein equation accordingly [10]. In this context, the dimension effect in this basic transport D/µ equation will be expected while the same system goes from bulk (3D) to nanoscale ranges (2, 1 and 0D). One such paradigm might have the cooperative behavior of temperature and electronic contributions, which can pertain to linear or nonlinear physics in diffusion-mobility equation.

Moreover, research in two dimensional (2D) materials has attracted much attention due to unusual physical and mechanical properties and also motivates to design the high performance devices [24, 25]. Noteworthy, the DOS of 2D materials does not dependent on energy and has less disorder effect at which one can ease to controls the energy dissipation in electronic and energy devices. However, the real quantum devices associate with the disorder as well as various typical interactions which lead to inhomogeneous electronic dispersion [26-28]. This dispersion is a direct evidence of entropy contribution to the systems at finite temperature. Usually high performance devices are fabricated using degenerate materials with high charge density. In the domain of many-electron interactions, the electrodynamics follows the collective behavior rule (like, hydrodynamics) and it favors the diffusion transport [29]. This continuum interaction in the many-body systems arises from the fast electron-electron scattering time ($\tau_{e-e} \to 0$) and it in turn gives rise to electronic transport as diffusion model rather than effective mass approximation of Drude model [21, 29-31]. The effect of total interactions on electronic transport can be determined by electronic compressibility (or DOS), which provides the direct evidence of diffusive conductivity in quantum materials (e.g., graphene) [21]. Here, the lower value of compressibility responsible for the extended localized states or incompressible band, and stronger compressibility is termed as the confined nondegenerate states [21, 22]. The first one manifests quantum flux, which is extracted from chemical potential jump and the later one relates with the electronic localization in which the charge transport can be thermally activated.

Based on the above ground, here we revisit the Einstein equation for ordered and disordered 2D degenerate materials (including organic semiconductors) through many-body physics. In this paper, we have developed the exact analytical solution for D/μ relation with the consideration of Gaussian DOS and Fermi-Dirac distribution for Schrödinger type materials. The Electronic DOS of Schrödinger materials are equivalent to that of nonrelativistic particles. The change of momentum (wave vector) variables due to the change of total interactions can be explained by entropy function. The collective behavior due to interactions must disturb the independent motion of fermions which leads to quasiparticles (fermions dressed by interactions) dynamics [32]. In this context, the expected diffusion and carrier density reduction can be evaluated by imperfect fermion entropy function, which elucidates the momentum dispersion in the interacting Fermi gas system. This is a direct consequence of disorder weightage on the charge redistribution and on the changes of diffusion transport. More detailed explanations of interactions effect via entropy are reported by many authors [29, 33-35]. In our model, the energy dispersion due to disorder, including thermal effect, is taken account through imperfect Fermi gas entropy term. Here, in such entropy modulated DOS and its consequences on diffusion-mobility properties are addressed at different thermodynamical limits, which effectively describes the temperature mediated incoherency in electronic states. Accordingly, rising density fluctuations along the consequential sites requires the statistical analysis on D/μ equation. This will be helpful to the experimentalists for thorough understanding of electronic fluidity behavior in high performance quantum devices at low temperature [29]. Moreover, we have demonstrated linear to nonlinear behavior of D/μ as a function of temperature and chemical potential, also our numerical results are verified by the experimental data of different quantum materials. Notably, the original Einstein equation is preserved within our extended Einstein Equations (derived for Schrödinger materials) in the limit of high temperature, $k_B T >> \eta$. The results suggest that our proposed paradigm suits for both the quantum and the classical systems (i.e., band to hopping descriptions) and importantly describes the coupled effect of electronic and thermal counterparts on D/μ relation.

This paper is organized as follows. In Section II, we present the derivations of carrier density, DOS (or electronic compressibility), and diffusion-mobility ratio for both generalized and degenerate 2D-Schrödinger materials. Also, the disorder effect on these parameters is described, through the imperfect fermion entropy term, which takes in to account the many-body interactions. Based on these formalism, the charge and energy perspectives on device performance (via electronic transport) for quantum materials at different physical domain are discussed in the Section III, using the data from earlier study [36]. In this connection, we are introduced the disorder effect on each parameters like, compressibility, diffusion, mobility and addressed the cooperative character of electronic and thermal contributions on device function. Also, we have shown the validity and importance of our extended Einstein D/μ - model in different thermodynamic limits. In Section IV, we have verified our extended Einstein model using earlier experimental and theoretical results with different degenerate materials like, graphene and $Bi_2Se_3$. Finally some important observations and is related implications are summarized in Section V.

## II. FORMALISM

### A. Charge density, electronic compressibility and D/μ - equation for Schrödinger materials

Generally, Einstein D/μ relation is defined as the carrier density times the inverse of electronic compressibility and it can be expressed as [5, 37],

$$\frac{D}{\mu} = \frac{n}{e(\partial n/\partial \eta)}, \qquad (1)$$

where *n* is the number density of carriers, *e* is the electronic charge, *η* is the chemical potential and $\frac{\partial n}{\partial \eta}$ is the electronic compressibility which is equivalent to that of electronic density of states [22, 38]. The real materials and devices are degenerate classes of high density limit, and hence the carrier density can be estimated using Fermi-Dirac distribution function ($f(E)$) and Gaussian DOS. Accordingly, the derived carrier density for 2D Schrödinger materials can be written as (see Appendix A),

$$n = \frac{2mk_BT}{\pi\sqrt{2\pi}\hbar^2}\frac{1}{\sigma}\ln\left(1+\exp\left(\frac{\eta}{k_BT}\right)\right)\left[1+\exp\left(-\frac{\varepsilon^2}{2\sigma^2}\right)\right], \quad (2)$$

where, $m$ is the carrier effective mass, T is the temperature, $\hbar$ is the reduced Planck constant, $\sigma$ is the normalized Gaussian variance $\sigma = \frac{\sigma_0}{k_BT}$ (or width), and $\varepsilon$ is the normalized energy $\varepsilon = \frac{E}{k_BT}$. Here, the carrier normalized energy can be described as (see A18),

$$\varepsilon = \frac{E}{k_BT} = 2\ln\left(1+\exp\left(\frac{\eta}{k_BT}\right)\right). \quad (3)$$

Inserting Eq. 3 into Eq. 2, we get the explicit form of charge density as,

$$n = \frac{2mk_BT}{\pi\hbar^2\sqrt{2\pi}}\frac{1}{\sigma}\ln\left(1+\exp\left(\frac{\eta}{k_BT}\right)\right)\left[1+\exp\left(-\frac{2\left(\ln\left(1+\exp\left(\frac{\eta}{k_BT}\right)\right)\right)^2}{\sigma^2}\right)\right] \quad (4)$$

The derivative of charge density with respect to the chemical potential is termed as the electronic compressibility. Using Eq. 4, we have derived the generalized compressibility expression as (see Appendix A),

$$\frac{\partial n}{\partial \eta} = \frac{2m}{\pi\hbar^2\sqrt{2\pi}}\frac{1}{\sigma}\frac{\exp\left(\frac{\eta}{k_BT}\right)}{\left(1+\exp\left(\frac{\eta}{k_BT}\right)\right)}\left\{\left[1+\exp\left(-\frac{2\left(\ln\left(1+\exp\left(\frac{\eta}{k_BT}\right)\right)\right)^2}{\sigma^2}\right)\right] - \frac{4\left(\ln\left(1+\exp\left(\frac{\eta}{k_BT}\right)\right)\right)^2}{\sigma^2}\exp\left[-\frac{2\left(\ln\left(1+\exp\left(\frac{\eta}{k_BT}\right)\right)\right)^2}{\sigma^2}\right]\right\} \quad (5)$$

The above Eq. 5 provides the electronic DOS information, which depends on temperature, effective chemical potential and Gaussian variance. Using compressibility expression, one can estimate the effective mass of a particle (electron/hole) which is basically originated with the many-body interaction. Substituting Eq. 4 and 5 into Eq. 1, the general form of diffusion-mobility relation (D/µ) for 2D Schrödinger materials can be formulated as,

$$\frac{D}{\mu} = \frac{k_BT}{e}\left[\frac{\left(1+\exp\left(\frac{\eta}{k_BT}\right)\right)\ln\left(1+\exp\left(\frac{\eta}{k_BT}\right)\right)}{\exp\left(\frac{\eta}{k_BT}\right)}\right]\left\{\frac{\left[1+\exp\left(-\frac{2\left(\ln\left(1+\exp\left(\frac{\eta}{k_BT}\right)\right)\right)^2}{\sigma^2}\right)\right]}{\left[1+\exp\left(-\frac{2\left(\ln\left(1+\exp\left(\frac{\eta}{k_BT}\right)\right)\right)^2}{\sigma^2}\right)\right]-\frac{4\left(\ln\left(1+\exp\left(\frac{\eta}{k_BT}\right)\right)\right)^2}{\sigma^2}\exp\left[-\frac{2\left(\ln\left(1+\exp\left(\frac{\eta}{k_BT}\right)\right)\right)^2}{\sigma^2}\right]}\right\} \quad (6)$$

This is the general D/µ expression for both the degenerate and the nondegenerate 2D materials. The Gaussian variance ($\sigma$) is basically the disorder width parameter and is related to the existence of degeneracy levels under the circumstances of applied electric field or magnetic field [21, 22]. Here, the formation of many-body incompressible electronic states (degeneracy levels) along with the landau level gap can be measured by the disorder width ($\sigma$) and the shift of chemical potential ($\eta$).

In zero dispersion (too weak Gaussian disorder), $\sigma \to 0$, the above general form of D/µ (Eq.6) becomes,

$$\frac{D}{\mu} = \frac{k_BT}{e}\left[\frac{\left(1+\exp\left(\frac{\eta}{k_BT}\right)\right)\ln\left(1+\exp\left(\frac{\eta}{k_BT}\right)\right)}{\exp\left(\frac{\eta}{k_BT}\right)}\right] \quad (7)$$

For degenerate cases of high density limit ($\eta \gg k_BT$), the Eq. (4), (5) and Eq. (6) are reduced as,

$$n = \frac{2m\eta}{\pi\hbar^2\sqrt{2\pi}}\frac{1}{\sigma}\left[1+\exp\left(-\frac{2\eta^2}{\sigma^2k_B^2T^2}\right)\right] \quad (8)$$

$$\frac{\partial n}{\partial \eta} = \frac{2m}{\pi\hbar^2\sqrt{2\pi}}\frac{1}{\sigma}\left\{\left[1+\exp\left(-\frac{2\eta^2}{\sigma^2k_B^2T^2}\right)\right]-\frac{4\eta^2}{\sigma^2k_B^2T^2}\exp\left[-\frac{2\eta^2}{\sigma^2k_B^2T^2}\right]\right\} \quad (9)$$

$$\frac{D}{\mu} = \frac{\eta}{e} \left\{ \frac{\left[1 + exp\left(-\frac{2\eta^2}{\sigma^2 k_B^2 T^2}\right)\right]}{\left[1 + exp\left(-\frac{2\eta^2}{\sigma^2 k_B^2 T^2}\right)\right] - \left(\frac{2\eta}{\sigma k_B T}\right)^2 exp\left(-\frac{2\eta^2}{\sigma^2 k_B^2 T^2}\right)} \right\} \quad (10)$$

For very weak disorder ($\sigma \rightarrow 0$), or zero disorder width ($\sigma = 0$) in degenerate materials, the D/μ equation becomes,

$$\frac{D}{\mu} = \frac{\eta}{e} \quad (11)$$

The Eq. (11) is the fundamental transport equation for quantum materials and it works very well at very low temperature regime. The above quantum diffusion-mobility relation is purely linear dependent on the chemical potential (or Fermi energy at zero temperature).

At high temperature domain of $k_B T \gg \eta$ (nondegenerate condition), the Eq. 7 is further reduced to the classical Einstein D/μ equation as,

$$\frac{D}{\mu} = \frac{k_B T}{e} \quad (12)$$

The above equation is the original Einstein equation, which is linearly depending on only the temperature.

### B. Entropy modulated charge density and its consequences on electronic compressibility and on D/μ - equation for Schrödinger materials

According to earlier models and reports [11, 39-42], the charge density and diffusion are limited by the thermal disorder; it can be quantified by the term of entropy. The electronic dispersion due to entropy alters the energy landscape in the materials (in asymmetry way), at which charge-energy flux is limited. To drag the entropy weightage on carrier flux changes, we have developed the entropy dependent charge density equation for common 2D systems as (see Appendix B),

$$n_S = n \, exp\left(-\frac{S}{2k_B}\right), \quad (13)$$

where, $n$ and $S$ are the carrier density in the absence of entropy (or thermal disorder) and entropy of a system, respectively. In this paper, the derived entropy expression can be written as (see Appendix C),

$$S = \frac{\pi^3}{12} k_B \frac{1}{\ln\left(1 + exp\left(\frac{\eta}{k_B T}\right)\right)} \quad (14)$$

In quantum limit of $\eta \gg k_B T$, the entropy formula will be simplified as,

$$S = \frac{\pi^3}{12} k_B \frac{k_B T}{\eta} \quad (15)$$

Using Eq. 1 and 13, the entropy modulated D/μ equation can be defined for 2D systems as,

$$\left(\frac{D}{\mu}\right)_S = \frac{n_S}{e\left(\frac{\partial n_S}{\partial \eta}\right)} = \frac{n}{e\left[\frac{\partial n}{\partial \eta} - \frac{1}{2k_B} n\left(\frac{\partial S}{\partial \eta}\right)\right]} \quad (16)$$

For such disordered 2D systems, the entropy modulated Gaussian carrier density equation can be written as,

$$n_S = \frac{2m k_B T}{\pi \sqrt{2\pi} \hbar^2} \frac{1}{\sigma} \ln\left(1 + exp\left(\frac{\eta}{k_B T}\right)\right) exp\left(-\frac{S}{2k_B}\right) \left[1 + exp\left(-\frac{\varepsilon_S^2}{2\sigma^2}\right)\right] \quad (17)$$

Here, the implicit form of entropically controlled normalized carrier energy expression as (see Eq. (D3)),

$$\varepsilon_S = 2 \ln\left(1 + exp\left(\frac{\eta}{k_B T}\right)\right) exp\left[-\frac{\pi^3}{24 \ln\left(1 + exp\left(\frac{\eta}{k_B T}\right)\right)}\right] \quad (18)$$

Applying Eq. 14 and 18 into Eq. 17, we get the final form of entropy modulated charge density equation as (see Appendix D),

$$n_S = \frac{2m k_B T}{\pi \hbar^2 \sqrt{2\pi}} \frac{1}{\sigma} \ln\left(1 + exp\left(\frac{\eta}{k_B T}\right)\right) exp\left[-\frac{\pi^3}{24 \ln\left(1 + exp\left(\frac{\eta}{k_B T}\right)\right)}\right] \left\{1 + exp\left[-\frac{2\left[\ln\left(1 + exp\left(\frac{\eta}{k_B T}\right)\right)\right]^2 exp\left(-\frac{\pi^3}{12 \ln\left(1 + exp\left(\frac{\eta}{k_B T}\right)\right)}\right)}{\sigma^2}\right]\right\} \quad (19)$$

Thus, the consequential effect by entropy on electronic compressibility can be described as,

$$\frac{\partial n_S}{\partial \eta} = \frac{2m}{\pi \hbar^2 \sqrt{2\pi}} \frac{1}{\sigma} \frac{exp\left(\frac{\eta}{k_BT}\right)}{\left(1+exp\left(\frac{\eta}{k_BT}\right)\right)} exp\left(-\frac{\pi^3}{24y}\right) \left\{ \left[1 + exp\left(-\frac{2y^2 exp\left(-\frac{\pi^3}{12y}\right)}{\sigma^2}\right)\right]\left(1+\frac{\pi^3}{24y}\right) - \left\{y\, exp\left(-\frac{\pi^3}{12y}\right) exp\left(-\frac{2y^2 exp\left(-\frac{\pi^3}{12y}\right)}{\sigma^2}\right)\left[\frac{4y}{\sigma^2}+\frac{\pi^3}{6}\right]\right\}\right\} \quad (20)$$

where, $y = ln\left(1 + exp\left(\frac{\eta}{k_BT}\right)\right)$.

Inserting Eq. 19 and 20 in to Eq. 16, the entropy modulated D/μ equation for 2D Schrödinger materials can be obtained as,

$$\left(\frac{D}{\mu}\right)_S = \frac{k_BT}{e}\left[\frac{y\left(1+exp\left(\frac{\eta}{k_BT}\right)\right)}{exp\left(\frac{\eta}{k_BT}\right)}\right]\left\{\frac{\left[1+exp\left(-\frac{2y^2 exp\left(-\frac{\pi^3}{12y}\right)}{\sigma^2}\right)\right]}{\left[1+exp\left(-\frac{2y^2 exp\left(-\frac{\pi^3}{12y}\right)}{\sigma^2}\right)\right]\left(1+\frac{\pi^3}{24y}\right)-\left\{y\, exp\left(-\frac{\pi^3}{12y}\right) exp\left(-\frac{2y^2 exp\left(-\frac{\pi^3}{12y}\right)}{\sigma^2}\right)\left[\frac{4y}{\sigma^2}+\frac{\pi^3}{6}\right]\right\}}\right\} \quad (21)$$

In the case of σ→0 (negligible or zero Gaussian width), the above D/μ equation is reduced as,

$$\left(\frac{D}{\mu}\right)_S = \frac{k_BT}{e}\left[\frac{\left(1+exp\left(\frac{\eta}{k_BT}\right)\right) ln\left(1+exp\left(\frac{\eta}{k_BT}\right)\right)}{exp\left(\frac{\eta}{k_BT}\right)}\right]\left(\frac{1}{1+\frac{\pi^3}{24\, ln\left(1+exp\left(\frac{\eta}{k_BT}\right)\right)}}\right) \quad (22)$$

In the degenerate cases of high density limit, $\eta \gg k_BT$, the Eq. 19, 20 and 21 can be revised as,

$$n_S = \frac{2m\eta}{\pi\hbar^2\sqrt{2\pi}}\frac{1}{\sigma} exp\left(-\frac{\pi^3 k_BT}{24\eta}\right)\left\{1 + exp\left[-\frac{2\eta^2 exp\left(-\frac{\pi^3 k_BT}{12\eta}\right)}{\sigma^2 k_B^2 T^2}\right]\right\} \quad (23)$$

$$\frac{\partial n_S}{\partial \eta} = \frac{2m}{\pi\hbar^2\sqrt{2\pi}}\frac{1}{\sigma} exp\left(-\frac{\pi^3 k_BT}{24\eta}\right)\left\{\left[1+exp\left(-\frac{2\eta^2 exp\left(\frac{\pi^3 k_BT}{12\eta}\right)}{\sigma^2 k_B^2 T^2}\right)\right]\left(1+\frac{\pi^3 k_BT}{24\eta}\right)-\left\{\frac{\eta}{k_BT} exp\left(-\frac{\pi^3 k_BT}{12\eta}\right) exp\left(-\frac{2\eta^2 exp\left(\frac{\pi^3 k_BT}{12\eta}\right)}{\sigma^2 k_B^2 T^2}\right)\left[\frac{4\eta}{\sigma^2 k_BT}+\frac{\pi^3}{6}\right]\right\}\right\} \quad (24)$$

$$\left(\frac{D}{\mu}\right)_S = \frac{\eta}{e}\left\{\frac{1+exp\left(-\frac{2\eta^2 exp\left(-\frac{\pi^3 k_BT}{12\eta}\right)}{\sigma^2 k_B^2 T^2}\right)}{\left[1+exp\left(-\frac{2\eta^2 exp\left(\frac{\pi^3 k_BT}{12\eta}\right)}{\sigma^2 k_B^2 T^2}\right)\right]\left(1+\frac{\pi^3 k_BT}{24\eta}\right)-\left\{\frac{\eta}{k_BT} exp\left(-\frac{\pi^3 k_BT}{12\eta}\right) exp\left(-\frac{2\eta^2 exp\left(\frac{\pi^3 k_BT}{12\eta}\right)}{\sigma^2 k_B^2 T^2}\right)\left[\frac{4\eta}{\sigma^2 k_BT}+\frac{\pi^3}{6}\right]\right\}}\right\} \quad (25)$$

For zero Gaussian width (or σ→0), the disorder (or entropy) limited D/μ equation is further reduced as,

$$\left(\frac{D}{\mu}\right)_S = \frac{\eta}{e}\left[\frac{1}{1+\frac{\pi^3 k_BT}{24\eta}}\right] \quad (26)$$

In pure quantum limit, $T\to 0$,

$$\left(\frac{D}{\mu}\right)_{T\to 0} = \frac{\eta}{e} \equiv \frac{E_F}{e} = \frac{mv_F^2}{2e} \quad (27)$$

Now, this relation preserves the earlier D/μ relation (see Eq. 11). In such limit, the diffusion-mobility linearly depends on only the parameter chemical potential. Here, D/μ basically provides one to one correspondence between the electronic information and the transport mechanism of a particular system. This equation is valid for all 2D Schrödinger type materials (bi and tri layer graphene and $MoS_2$, etc.)

### C. Entropy effect on diffusion and on mobility calculation in Schrödinger systems

The diffusion limited by thermal disorder in the 2D electronic systems can be expressed as (see Appendix B),

$$D_S = D\, exp\left(-\frac{S}{4k_B}\right) = D\, exp\left(-\frac{\pi^3}{48\, ln\left(1+exp\left(\frac{\eta}{k_BT}\right)\right)}\right) \quad (28)$$

For degenerate conditions, the above Eq. 28 is reduced as,

$$D_S = D(\eta,T) = D\, exp\left(-\frac{\pi^3 k_B T}{48\eta}\right) \quad (29)$$

The explicit form of entropy contribution on mobility is described as (see Appendix E),

$$\mu_S = \mu(\eta,T) = \mu\left[1 + \frac{\pi^3}{24\, ln\left(1+exp\left(\frac{\eta}{k_B T}\right)\right)}\right] exp\left(-\frac{\pi^3}{48\, ln\left(1+exp\left(\frac{\eta}{k_B T}\right)\right)}\right) \quad (31)$$

At high charge density limit, the Eq. (31) becomes,

$$\mu_S = \mu(\eta,T) = \mu\left[1 + \frac{\pi^3 k_B T}{24\eta}\right] exp\left(-\frac{\pi^3 k_B T}{48\eta}\right) \quad (32)$$

Here, $\mu = \frac{e}{k_B T} \frac{exp\left(\frac{\eta}{k_B T}\right)}{\left(1+exp\left(\frac{\eta}{k_B T}\right)\right) ln\left(1+exp\left(\frac{\eta}{k_B T}\right)\right)} D$, and $D$ is the diffusion coefficient in the absence of entropy effect, respectively. For degenerate condition, $\mu = \frac{e}{\eta} D$.

$$\mu_S = \mu\left[1 + \frac{S}{2k_B}\right] exp\left(-\frac{S}{4k_B}\right) \quad (30)$$

Thus, the thermodynamically parameterized disorder state-mobility equation can be written as,

## III. RESULTS AND DISCUSSION

Using our formalism, we have investigated the diffusion-mobility transport mechanism using carrier density and compressibility of some 2D quantum Schrödinger materials at different thermodynamical limits. To this calculation, we have used some of the earlier reported experimental as well as theoretical data for the analysis.

### A. Diffusion-mobility transport in Schrödinger materials (General form)

The multi-layer graphene, molybdenum disulfide and layered organic films are the best examples for Schrödinger materials. In these materials, the carrier motion follows the non-relativistic dynamics. Apparently, many authentic studies explain the formation of electron-hole puddles in 2D quantum materials and its dependency on disorder. In such cases, the existence of inhomogeneous charge density expedites the diffusion transport [21, 36]. This is mainly related with the shape of DOS as well as effective interactions on the particle, which can be analyzed by chemical potential. Generally chemical potential depends upon the gate-voltage, carrier doping and the applied magnetic field, etc. [22, 43, 44]. As reported from earlier study [36], through electron-hole puddles the calculated electron effective mass in bilayer and trilayer graphene are $0.063 m_e$ and $0.082 m_e$, respectively. We have used these values in Eq. 4 and 5 to measure the carrier density and the compressibility (or DOS). Calculated carrier density and DOS for wide range of chemical potential with different Gaussian variance at different temperatures are plotted, which are shown in Fig. A1 and A2, respectively. The negative chemical potential values are commonly referred as the electron localization domain, in which the carrier density can be activated by thermal energy for electronic transport (see Fig. A1). For disordered systems, the potential energy landscape will be shallow or in deep depth, which depends on the disorder values. The trapped sites are measurable by the parameter of negative chemical potential and its differences. In principle, the presence of potential minima of the energy landscape act as trap sites and it shut the diffusion transport, but it can be activated by temperature which is noted in Fig. A1. For high temperature, the activated charge carrier values are more in the localization (or negative chemical potential) domain. Generally, the positive region of chemical potential is the delocalized carrier transport region. Now, the diffusion-mobility is directly proportional to the chemical potential. The energy landscape (including shape) and its width are fixed by the parameter Gaussian variance (σ). In this way, the carrier density variation with respect to the chemical potential gives rise to the electronic compressibility, which describes the compressible or incompressible nature of electronic states, which is equivalent to that of DOS of the materials. The Fig. A2 shows the chemical potential dependent DOS at different temperatures for different Gaussian width. At very low temperature ($T\to 0$), the DOS is negligible in the wide range of negative chemical potential values and it takes sharp peak at nearer to zero chemical potential, then DOS follows constantly larger value in the entire positive side of chemical potential. In this limit ($T\to 0$), DOS behaves like a step function for different Gaussian variance. For high temperature, DOS exponentially increases with the chemical potential (or energy) when Gaussian variance approaches to zero, σ→0 (see Fig. A2). At the same time, shape of DOS is Gaussian in the larger value of σ. It is to be noted that the peak and width of DOS are determined by σ and T. Based on the above analysis, one can predict the carrier contribution to the device performance through above parameters η, T and σ.

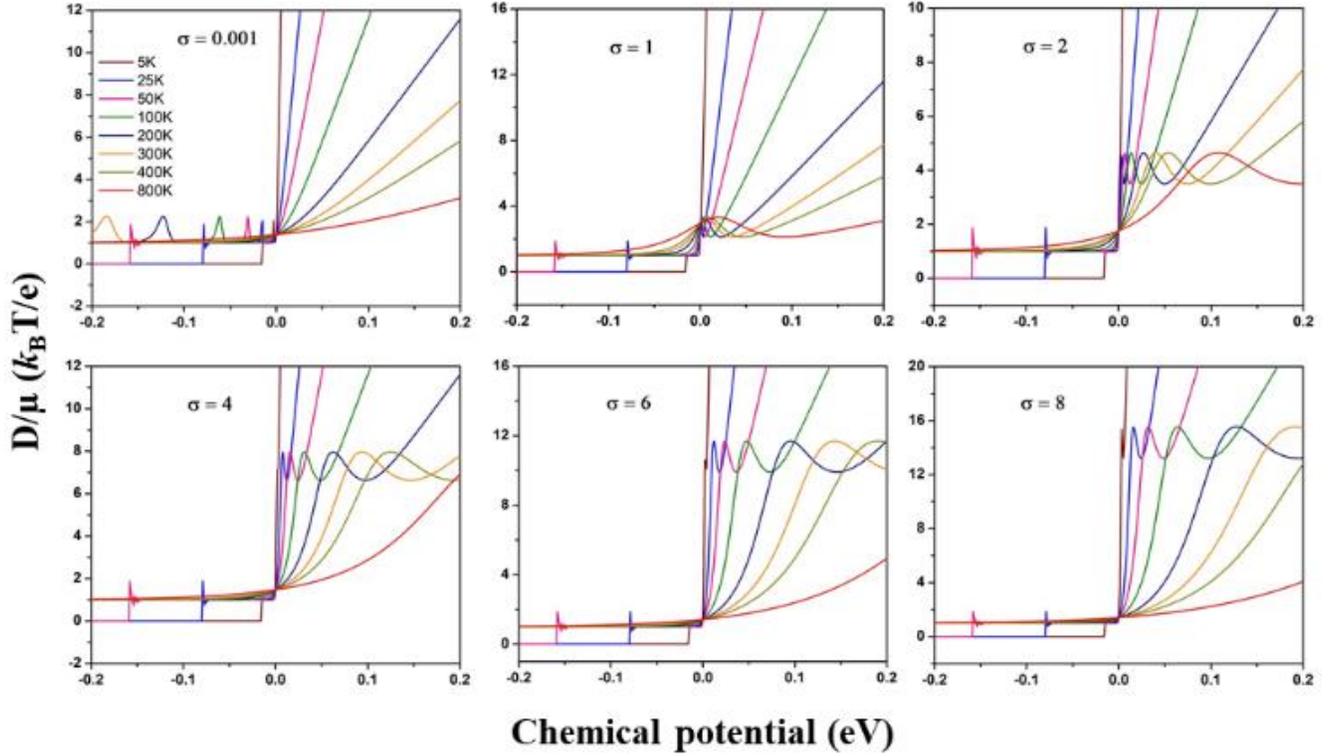

FIG. 1. Enhancement of diffusion- mobility ratio with respect to the chemical potential at different temperature values for different Gaussian invariance. Validity and limitations of Einstein relation depends on chemical potential of the system at a given temperature. The plot shows the validity of Einstein relation in wide chemical potential ranges for high temperature regime, and deviates in degenerate limit of low temperature, $\eta >> k_B T$.

In this work, the calculated D/μ factor for wide ranges of η and of T at different σ values is shown in Fig. 1. Here the ratio between carrier density and DOS provides the D/μ factor (see Eq. 1). We interestingly find that there is the absence of diffusion-mobility transport in the whole negative chemical potential ranges in the zero temperature limits. While increasing the temperature, the diffusion phenomena are observed even in localization domain of negative chemical potential region (see Fig. 1). Here, the activated diffusion transport occurs which is mainly due to the thermal energy, termed as thermally activated diffusion-mobility which is the classical Einstein relation. At each finite temperature, the validity and limitations of Einstein D/μ can be predicted at different range of chemical potential for different Gaussian variance. Too weak disorder, the Gaussian variance might be σ → 0. For example, at 5 K in the Gaussian variance of σ = 0.001 cases, the absence of diffusion (D/μ = 0) is noted up to the chemical potential values of -16 meV, (η ≤ -16 meV). The validity of Einstein equation is observed in the range of -16 < η < 0. At zero chemical potential in 5 K, the value of D/μ is 1.3863 times $k_B T/e$. The enhancement of D/μ is observed for positive values of whole chemical potential (η ≥ 0), $\frac{D}{\mu} > \frac{k_B T}{e}$. In the similar way, at 50 K for σ = 0.001, the calculated diffusion is zero up to the chemical potential value of -159 meV, (D/μ = 0; η ≤ -159 meV). Also, the validity of Einstein equation is observed in the range of -158 to -7 meV, $\frac{D}{\mu} = \frac{k_B T}{e}$; $-158 \leq \eta \leq -7$ meV. Beyond that (chemical potential ranges η > -7 meV), the enhancement of D/μ is noted, i.e., $\frac{D}{\mu} > \frac{k_B T}{e}$. Importantly, the validity of original Einstein D/μ value ($k_B T/e$) has been noted in the high temperature values for vast chemical potential ranges, see Fig. 1. Absence of diffusion is mainly responsible for insulator characteristics of the given systems. As noted from Fig. 1, for very weak disorder (σ → 0) materials, the existence of Gaussian width lies in the negative chemical potential domain. While increasing the disorder width, the formation of Gaussian wave-packets shifts towards the positive chemical potential landscape. Due to the Gaussian shape DOS (see Fig. A2), the Gaussian-like transport is expected in the diffusion based mobility for 2D quantum materials.

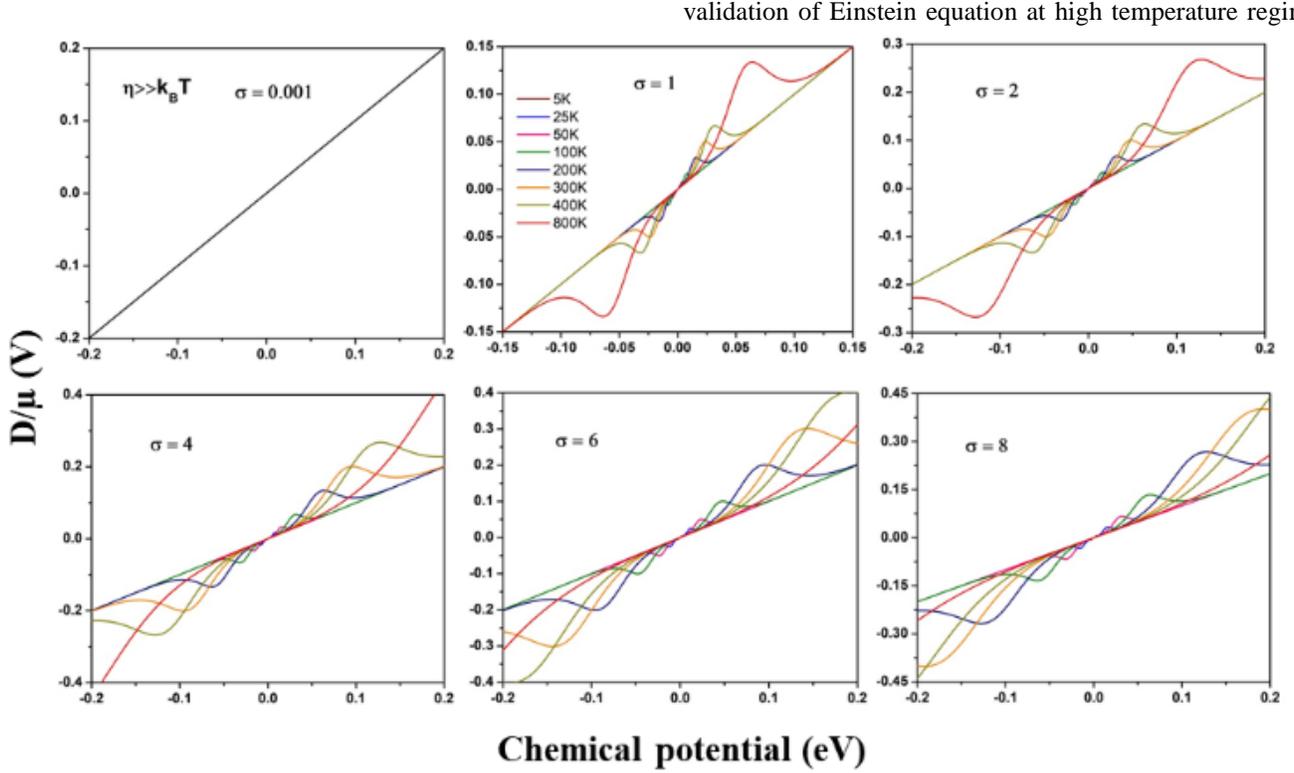

FIG. 2. Enhancement of D/μ as a function of chemical potential at different temperature values in different Gaussian variances for degenerate materials. The chemical potential jumps at different temperatures and Gaussian variances leads to quantum flux. The inversion symmetry is observed in electro-hole D/μ transport.

In principle, the D/μ factor is equivalent of potential, known as activation potential for such transport. When $k_B T \gg \eta$, the corresponding D/μ potential is referred as thermally activation potential for carrier motion. On the other hand, in degenerate limit ($\eta \gg k_B T$) the diffusion-mobility is enhanced by chemical potential. In this study, the survival time of carrier at each state can be defined using uncertainty relation, $t = \frac{\hbar \mu}{2eD}$. Accordingly calculated survival time at different temperatures for wide chemical potential ranges in various Gaussian widths is shown in Fig. A3. In the negative chemical potential regime (localized states), the carrier survival time is more at low temperature cases. The surveying time is decreasing with increasing the temperature which is directly observed from Fig. A3. Here, temperature activates the carrier motion along the consequential localized sites in the disordered materials. At high temperature values, there is no significant variation in survival time for different chemical potential (or carrier energy) values. For instance, at 800K in the Gaussian width σ = 8, the survival time follows a linear fashion, and also independent of chemical potential but depends only on temperature (see Fig.A3). Now the Einstein D/μ equation is valid. In this study, the survival time plot explicitly consolidates the validity and limitations of Einstein equation (see Fig. A3). The results clearly show the validation of Einstein equation at high temperature regime and deviation is observed in low temperature limit.

### B. Diffusion-mobility transport in Schrödinger materials (Degenerate form)

Practically, devices are configured by high charge density degenerate materials ($\eta \gg k_B T$) and its characteristic behavior can be modified via doping and by controlling the bias-voltage. In such degenerate limit, the carrier density, DOS and D/μ factor are calculated using Eq. 8, 9 and 10, respectively. In this case, the electronic contribution is more, rather than temperature effect on electronic transport most of the quantum materials, even at room temperature. Here, the chemical potential (or Fermi energy) is crucial for device performance. Because the carrier density in degenerate materials is directly proportional to the chemical potential; there is no direct thermal energy counterpart in it. For weak disorder cases σ → 0, the D/μ value linearly depends only on chemical potential and is shown in Fig. 2. At zero limit temperatures for ideal quantum 2D systems, the diffusion-mobility might be a perfect linear relationship with respect to chemical potential. In the context of σ → 0 and of T → 0, the carrier density linearly varies with chemical potential, and there is a fixed DOS (or electronic compressibility) in wide range of chemical potential (see Figs. A4 and A5).

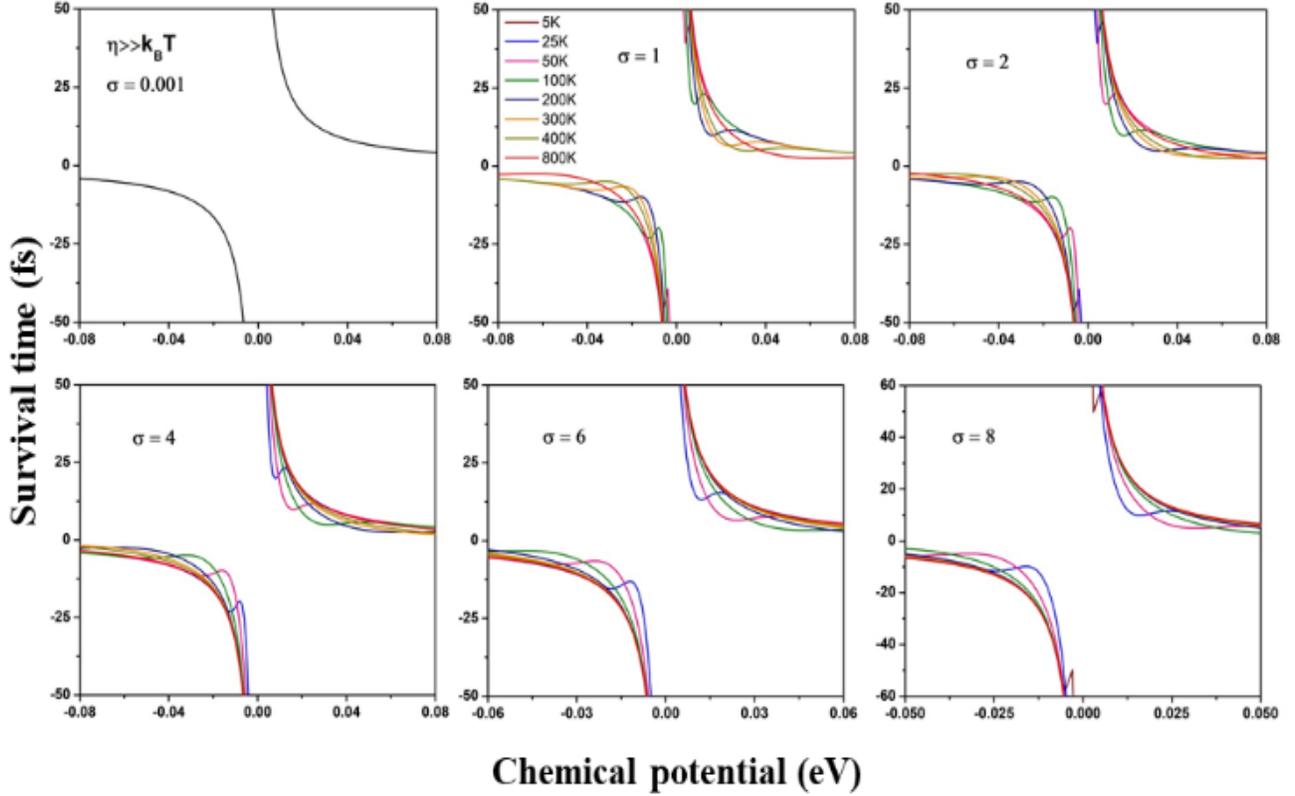

FIG. 3. Carrier survival time as a function of chemical potential at different temperature values in different Gaussian variances for degenerate materials. On the basis of quantum flux the survival time is varied with respect to the chemical potential. The plot shows the time reversal symmetry behavior in D/μ for the degenerate ideal 2D Schrödinger materials.

Here, the enhancement of D/μ factor is comparably so higher than original Einstein value of $k_BT/e$. Due to temperature and Gaussian width, the carrier density and DOS are modified; accordingly carrier transport in the form of diffusion-mobility is employed in such degenerate 2D materials. In the present study, the DOS follows the mirror symmetry behavior while chemical potential moves from positive region to negative region which is observed in Fig. A5. Moreover, here the electronic transport properties are equally solicited in both positive (for electron) and negative chemical potential (for hole) values, which turn out the inversion symmetry characteristics on electron-hole D/μ transport. Typically, if the applied bias voltage or electric fields equally modifies the electron and hole states (e.g., LUMO and HOMO) in which the calculated D/μ for both hole and electron are of same magnitude with opposite sign ($\eta_e = -\eta_h$), which is normally expected in ordered (or periodic) 2D materials. The inversion symmetrical nature of electron-hole D/μ equation can be directly noted from survival time versus chemical potential plot, Fig. 3. On the basis of temperature and Gaussian variance, the trend of carrier survival time can be analyzed.

In this extent (for pure ideal quantum 2D devices, i.e., zero dispersion σ → 0), the governed D/μ equation is equivalent to that of η/e (see Eq. 11). Accordingly, we can be redefined the Shockley diode current density equation as $J = J_0 \left[exp\left(\frac{eV}{\eta} - 1\right)\right]$, where, $J_0$ is the saturation current density and $V$ is the applied voltage. Our model clearly emphasis the quantum contribution to the D/μ and to the current density equation at very low temperature, generally follows the linear function of chemical potential.

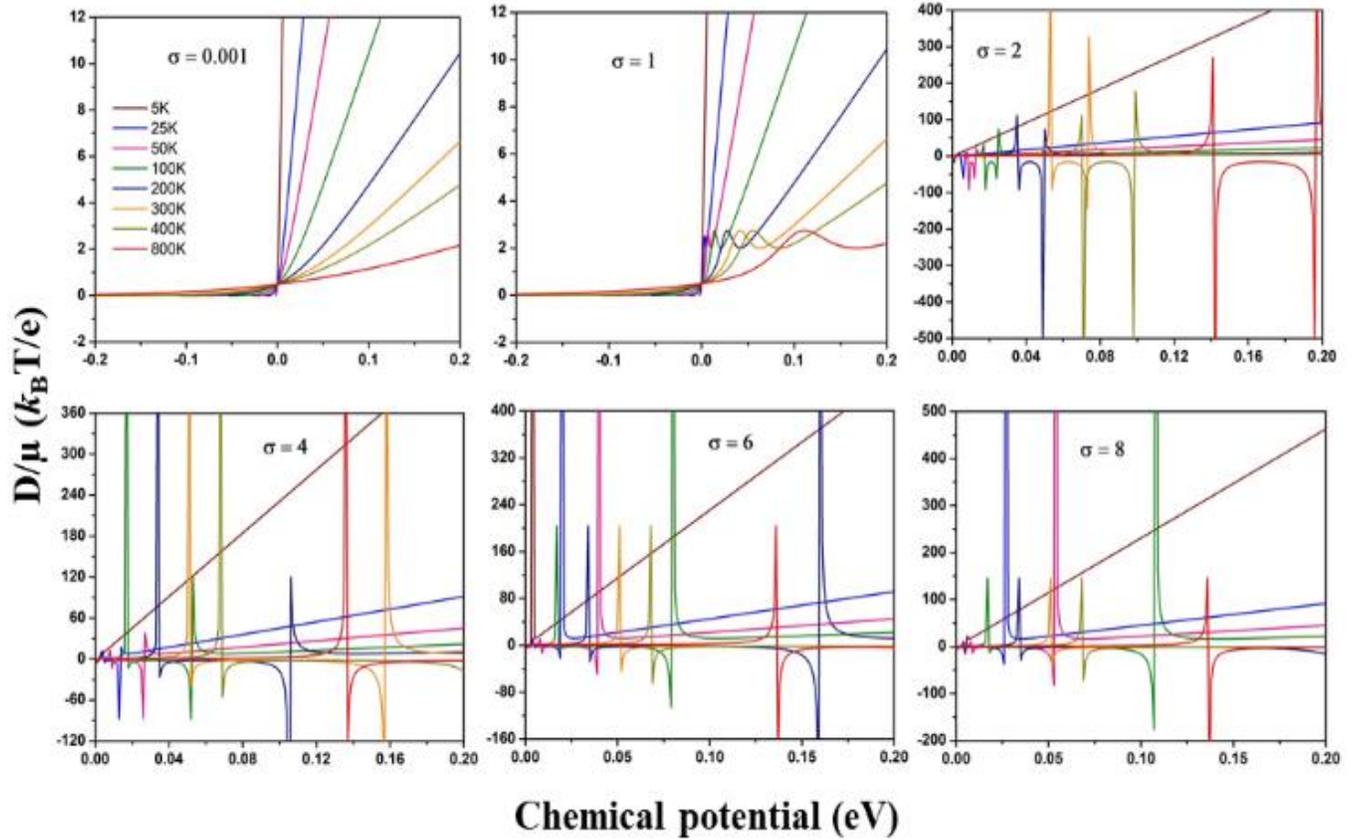

FIG. 4. Absence of diffusion-mobility is noted in the localized region of negative chemical potential, due to entropy effect. Enhancement of D/µ is started from peak for larger value of Gaussian variance in positive chemical potential side (delocalized regime). The carrier oscillation is expected in the sharp peak window.

### C. Entropy modulated diffusion-mobility transport in Schrödinger materials (General form)

The thermal and quantum fluctuation in N-particle system effectively modifies the eigen values of electronic states which can be analyzed by imperfect Femi-gas entropy method. In such a way, the occurred electronic dispersion in any 2D electronic systems is quantified by entropy term, which contains two simple parameters, temperature and chemical potential. Accordingly derived entropy modulated carrier density expression and its consequences on DOS and on D/µ factor gives rise to significant modification in the transport properties, which importantly deals the realistic challenges in device performance. The generalized diffusion-mobility equation does not explain the carrier motion in rough energy landscapes [45]. To overcome this issue, we have included entropy ($S$) along with Gaussian variance ($\sigma$) on D/µ derivation (see Eq. 21 and 25), which elucidates the carrier dynamics in rough (disordered) landscapes and also in perturbed regimes. It is to be noted that the presence of entropy, limits the diffusion (compare Fig. 1 and 4), agrees

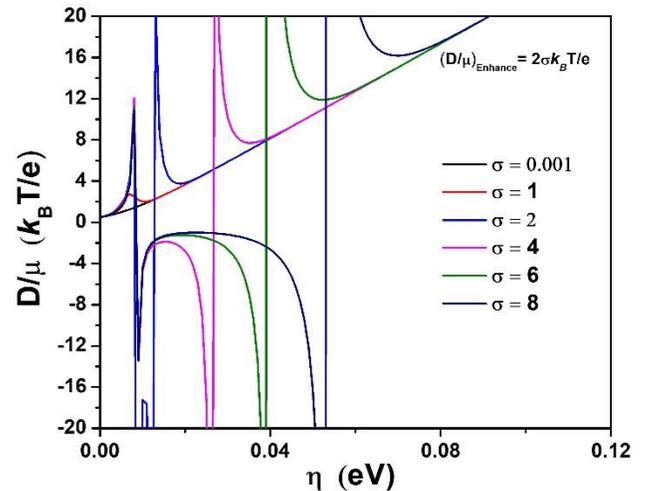

FIG. 5. Starting point values of D/µ enhancement at 50 K in the different Gaussian variance ($\sigma$). For example, at $\sigma = 6$, the origin of enhancement is $12 k_B T/e$.

with the diffusion limited by thermal disorder by Troisi *et al* [39, 41, 46].

The presence of entropy significantly suppresses the diffusion based mobility in the localization regime (negative chemical potential domain) for all temperatures at different Gaussian width. In this domain, the absence of diffusion turns into insulator behavior, which practically arises from various defects and carrier energy scattering of materials [47]. As noted in Fig. A7, the compressible DOS (or localized nature) and minimum probability of estimated mixed states highly resist the self-diffusion phenomena. Also, the carrier density contribution for transport is relatively very less which is shown in Fig. A6. It has been found that there is a D/μ enhancement with respect to the chemical potential in the positive domain (delocalized region). While the temperature moves from low to high values, the enhancement of D/μ value is apparently reduced (see Fig. 4) for different Gaussian variance. In this study, the Einstein transport $\left(\frac{D}{\mu} = \frac{k_B T}{e}\right)$ is preserved in the appropriate chemical potential at each finite temperature. Beyond certain *T* and η values, the D/μ takes the deviation from Einstein value, $\frac{D}{\mu} > \frac{k_B T}{e}$. For σ → 0, the smooth enhancement of D/μ is appeared. On other hand (for larger σ), the enhancement starts with double sharp peak and thereafter it smoothly increases with the chemical potential (see Fig. 4). Here, the presence of each peak underlies in the form of first order derivative of Dirac-delta function, $\frac{d}{d\eta}\delta(\eta)$. For instance, in the case of σ = 4 and T = 50 K, the 'D/μ versus η' plot originally starts from zero diffusion and takes two Dirac-delta derivative kind peaks with the finite gap and is end up with $\frac{D}{\mu} = 2\sigma \frac{k_B T}{e} = 8 \frac{k_B T}{e}$, which is plotted in Fig. 5. At this starting point of D/μ enhancement, the chemical potential value is nearly 36 meV. Thereafter the D/μ plot linearly varies with the chemical potential. The same trend is observed for all Gaussian width (σ) at different temperatures. For instance, the continuous D/μ enhancement started from $\frac{D}{\mu} = 16 \frac{k_B T}{e}$ in Y-axis at the chemical potential of 71 meV (X-axis) for σ = 8 and T = 50 K. Here the gap between two consequent peaks mainly depends on temperature. For low temperature, the gap is less but it increases with temperature. At T → 0 and at σ → 0, there is no peak and the D/μ enhancement directly starts from zero diffusion point, $\frac{D}{\mu} = 2\sigma \frac{k_B T}{e} = 0$. The peak sharpness and width are normally characterized by both T and σ.

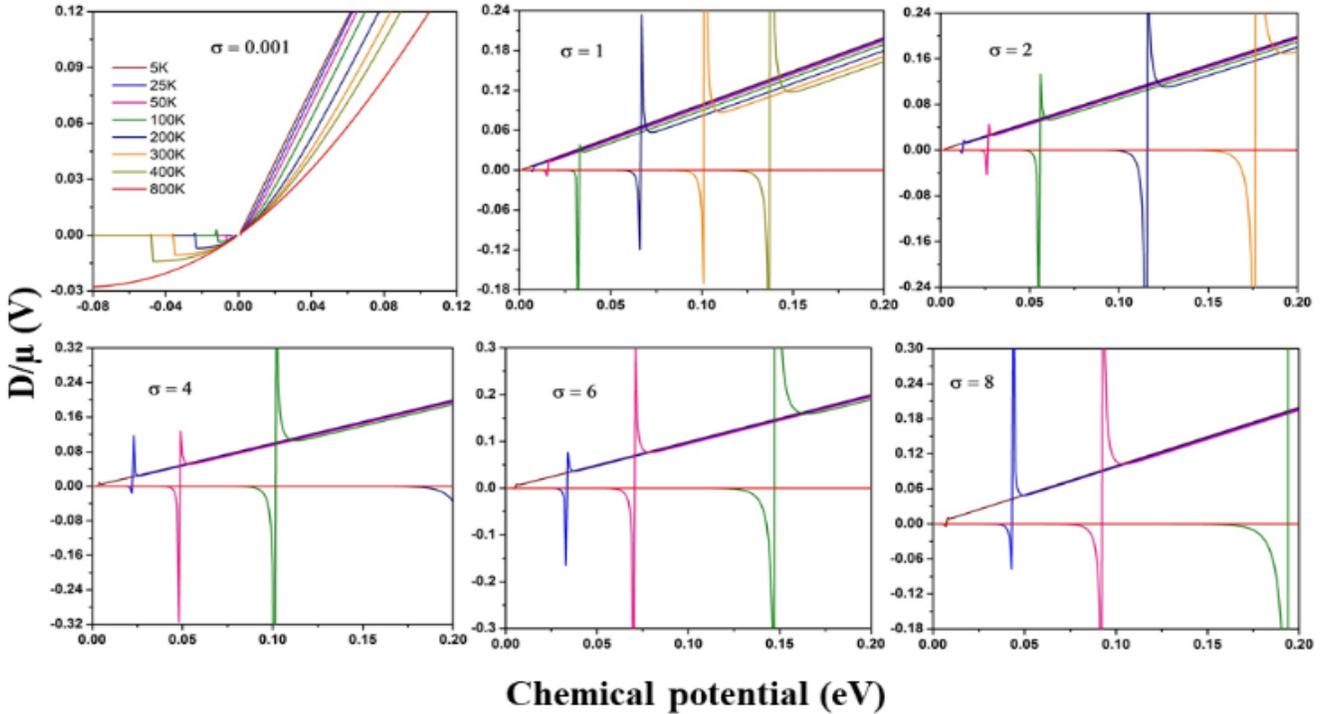

FIG. 6. Enhancement of entropy modulated D/μ values with respect to the chemical potential at different temperature for various Gaussian widths. The presence of entropy limits the carrier diffusion in the localized region (η < 0). The existence of peak indicates oscillate nature of carrier dynamics at finite temperature for different Gaussian variances. From that oscillation motion the D/μ enhancement started.

The existence of up and down peak in both the positive and the negative Y-axis (or in D/μ factor) with respect to the finite chemical potential (X-axis) causes the dipole moment. Based on number of up and down peak the multipole states arise. We interestingly find that there is a positive value of electronic compressibility (or DOS) in a weakly disorder limits (σ < 3), and negative compressibility is observed for large Gaussian disorder variance (σ) which is shown in Fig. A7. Based on DOS in each η at different T and σ, the D/μ factor has appeared. In the $\frac{d}{d\eta}\delta(\eta)$ region (or peak window) in D/μ plot, the carrier takes the up and down oscillation due to dipole effect, which can be defined by η and T at every σ. This peak width (in η units) strongly depends on T and σ, also to be termed as oscillation window. Now, the carrier activation is initiating with this jerk oscillated motion in such polaron states and it continuously increases the D/μ for different T. Here the perceived thermal fluctuation in D/μ-axis direct evidences of entropy effect on electronic transport. This oscillation nature of carrier dynamics is directly seen from survival time graph, Fig. A8. For σ →0, we do not find any time peak, and it smoothly decreases with the chemical potential. At low temperature, the survival time steeply decreases and a time delay occurs while temperature moves from low to high values, which is due to the thermally activated phonon scattering (also see ref. 36 and 37). It has been observed that there is a peak in survival time graph for higher order Gaussian variance which intriguingly tells the back-forth oscillated (or down-up) motion in the activation transport region (see Fig. A8), after that the dynamics turned-out in a straightforward translational manner. Here, the back-forth motion quantitatively connects with the nature of DOS, whether DOS has positive or negative values. In this study, the oscillation as well as the translational time window depends upon both the temperature and the Gaussian variance.

### D. Entropy modulated diffusion-mobility transport in Schrödinger materials (Degenerate form)

In this section, the entropy effect on D/μ value in degenerate quantum materials ($\eta >> k_B T$) is discussed. The microscopic understanding of electronic transport in such materials needs the knowledge of thermally coupled carrier density and its compressibility in electronic states, which connects with D/μ. Apparently thermal and quantum fluctuations in electronic states moderately deviates the transport properties through dispersion (unlike ideal transport systems, see Fig. A10 and 6), which can be quantified hereby in Eq. 23-25. As described in earlier section, the thermal fluctuation is the dominant factor for diffusion-mobility transport in generalized cases (see Fig. 4). To achieve good performance, the real devices are usually designing using degenerate materials. In practical condition, various typical interactions have to be included via appropriate thermodynamical quantities for thorough understanding in any physical devices. In this context, here we have addressed the effective interactions through the entropy term (derived from imperfect Fermi-gas model) on D/μ equation. The calculated entropy modulated carrier density and its consequences on DOS (or compressibility) and on D/μ are shown in Fig. A9, A10 and Fig. 6, respectively. Here the observed deviation in D/μ transport from ideal degenerate Schrödinger materials (compare Fig. 2 and 6), purely depends on many-body entropy function in Eq. 25. Due to applied electric and magnetic field, the existence of degeneracy in Gaussian-like DOS here is fixed by Gaussian variance which is associated with the disorder (see Eq. 24 and 25).

As noted in carrier density plot (see Fig. A9), the presence of entropy reduces the amount of charge density contribution to the electronic transport in localization region of negative potential values. On the other hand of delocalization domain, the D/μ transport steeply enhances with the chemical potential which is shown in Fig. 6. The enhancement will vary with the different Gaussian parameter for different temperature. In the zero dispersion limit (σ → 0), the diffusion based mobility is expected in the negative chemical potential regime, and this magnitude increases with the temperature (see Fig. 6). At the same time, the D/μ factor smoothly increases with chemical potential in the positive domain; also, here the slope value varies with respect to the temperature. For example, at σ = 0.001, the values of DOS end up at near zero value in the negative chemical potential region (see Fig. A10). But for σ = 1, 2, 4, 6 and 8 values, the DOS lift up towards the positive Y-axis (+ve DOS) and thereafter it traverses to positive chemical potential side (X-axis). In this extent, the appearance of DOS along the negative part of DOS (Y-axis) follows the parabolic shape with respect to the chemical potential (positive region). This traversing nature of DOS is responsible for sharp peak existence in D/μ plot for all T, at all different σ, which are shown in Fig. A10 and 6. Here, the peaks are looking like the derivative of Dirac-delta function, $\frac{d}{d\eta}\delta(\eta)$. Up to that peak point, there is absence of diffusion-mobility and thereafter the D/μ enhancement starts. This zero diffusion range is the insulator regime (in η units) which will be varying with respect to the temperature and the Gaussian variance (see Fig. 6). In the cases of T = 100 K and σ = 2, there is no diffusion-mobility up to the chemical potential value of around 50 meV, after that the peak appeared within the range of 51-61 meV (11 meV peak width). From this peak (from 62 meV), the D/μ enhancement starts and increases continuously with the chemical potential. At the same T = 100 K and for σ = 6, the absence of diffusion is observed up to the chemical potential at 128 meV, i.e., D/μ = 0 for η ≤ 128 meV. In this analysis, the peak arises from 129 meV and ends up at 165 meV and thus estimated oscillation width is 37 meV. Thereafter the D/µ continuously increases with the chemical potential. The results emphasis that the carrier dynamics starts with one jerk based back-forth (or down-up) oscillation and then it takes continues smooth D/µ enhancement along with the chemical potential. Within this small window (in η units), the carrier being in oscillated motion and its amplitude height and window width depends upon the chemical potential fluctuation which is termed as quantum fluctuation. From the above analysis, the oscillation window increases from 11 to 37 meV while the Gaussian variance changes from 2 to 6 at finite temperature of T = 100K. From these findings, the dipole moment and its oscillation window can be explicitly determined, which may be aligned by Gaussian variance with the aid of applied electric and magnetic field. The magnitude of chemical potential jumps (or quantum flux) can be analyzed by carrier traversing between the Landau levels (LLs) which is explained in various earlier studies [22, 38, 44]. It is to be conclude that the quantum flux is associated with the entropy modulated degenerate materials, and thermal flux associated with entropy modulated generalized (e.g., non-degenerate) materials (compare Fig. 6 and Fig. 4). The entropy modulated D/µ Eq. 25 for a degenerate material is reduced to Eq. 10, while the temperature effect is zero. We importantly note that there is symmetry breaking rule in D/µ transport due to entropy contribution on it (compare Fig. 2 and 6). Specifically the presence of entropy causes the dissimilar DOS with respect to the chemical potential values at different T and σ. In this extent, the nature of entropy modulated D/µ transport in degenerate systems at different thermodynamical conditions (in terms of T, σ and η) is directly measurable from survival time graph which is plotted in Fig. A11. It is noteworthy that, the activation chemical potential region as well as the traversing chemical potential ranges in this D/µ transport can be directly monitored. For instance, in the case of T = 200K and σ =1, the transport activation takes place at around 60 meV (see Fig. A11 and Fig. 6). Likewise, the expected activation chemical potential is around 52 meV for σ = 2 at T = 100 K, which is clearly noted in Fig. A11.

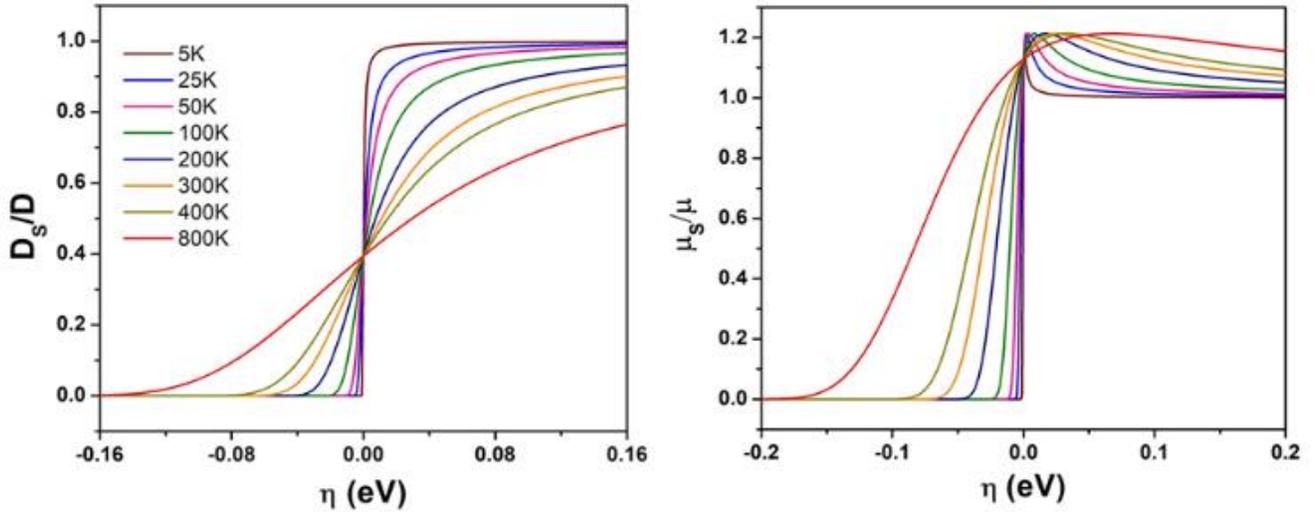

FIG. 7. Entropy effect on diffusion and on mobility is plotted for different temperature values. Thermally activated transport is observed in localized domain (η < 0). Thermally limited diffusion is occurred in the delocalized regime (η > 0), but mobility moderately increases with the temperature. Through appropriate entropy (with the aid of temperature and chemical potential), one can achieve the good transporting ability, at which the entropy mobility ($\mu_S$) relatively larger than the ideal mobility ($\mu$).

### E. Entropy effect on diffusion and on mobility

In earlier sections C and D, we have addressed the entropy effect on D/µ factor for both generalized and degenerate cases. To understand the thermally coupled electronic disorder effect individually on diffusion and on mobility, here we have derived the entropy dependent diffusion ($D_S$) and mobility ($\mu_S$) equations, on the basis of the zero dispersion (σ = 0) assumption. In the present study, the entropically modified diffusion and mobility are calculated at different temperatures in wide range of chemical potential values using Eq. 28 and 31, respectively, which is plotted in Fig. 7. In localization limit η < 0, the diffusion is almost zero at low temperature. The presence of disorder normally leads to shallow potential in the energy landscape of the materials. This shallow depth is measurable by entropy term, and quantifies the localization strength. In strongly localization limit of η = -∞ [48], there is no diffusion phenomena which is generally termed as insulator regime. As observed in Fig. 7, the diffusion is thermally activated while the temperature moves from lower to higher values. Similar trend has been noted in mobility plot for localized regime (or for negative chemical potential region). At low temperature T = 5 K, the absence of diffusion noted up to the chemical potential value of -2 meV. Thereafter, the diffusion increases very steeply within small chemical potential window and finally reaches the equivalent of ideal diffusion limit (i.e., zero entropy effect). At temperatures 100, 200 and 300 K, the predicted zero diffusion ranges are -15, -31 and -47 meV, respectively. Similar tendency is observed in mobility calculation. In this study, both the diffusion and mobility transport start (from zero value) from the chemical potential values of around -16, -32 and -48 meV in the temperatures of 100, 200 and 300K respectively. This study clearly emphasis that the localized carrier can be activated by temperature in the disordered systems. Moreover in the delocalized carrier region ($\eta > 0$), the studied carrier diffusion is significantly affected by temperature (see Fig. 7), due to the thermally coupled scattering effect. In the diffusion analysis, we mainly summarized two corollaries; (1) the diffusion is thermally activated in the localized regime ($\eta < 0$), (2) diffusion is limited by temperature in delocalized cases $\eta > 0$, due to thermal disorder. These findings are in agreement with the earlier studies [36, 41, 42, 46].

Besides that, the mobility in localized regime can be activated by temperature which is noted in Fig. 7. Here, we observe that the thermally activated mobility values are comparatively larger than the thermally activated diffusion transport in each chemical potential values, which can be analyzed from slope values (see Fig. 7). Also, we interestingly observed that at low temperature there is larger value of disordered mobility ($\mu_S$) than the ideal mobility ($\mu$) around the charge neutrality point (CNP), $\eta = 0$. For example, at T = 5 K, the disordered mobility reaches maximum ($\mu_S = 1.175\mu$) at $\eta = 1$ meV, thereafter this mobility suddenly reduced towards $\mu_S = \mu$ and then it follows the same ideal mobility value for all chemical potential values. At room temperature, $\mu_S$ reaches the maximum of $1.2131\mu$ at $\eta = 25$ meV, thereafter it reaches the ideal mobility ($\mu$) and it continues the same $\mu$ value for all $\eta$ values. Based on the above analysis, it has been concluded that the peak of $\mu_S$ changes (or shifts) with respect to the chemical potential for different temperatures (from CNP to high $\eta$ values, see Fig.). For delocalized band transport regime ($\eta > 0$), the mobility relatively decreases while the temperature moves from high to low values. Importantly, we conclude that the mobility is enhanced at finite entropy value, which can be achieved with the appropriate temperature and chemical potential.

## IV. ANALYSIS AND APPLICATIONS

Various experimental studies confirm the deviation in Einstein relation, which is commonly estimated by diode ideality factor (or enhancement factor) [4, 8, 9, 13]. Importantly, at low temperature, the Einstein relation of D/µ fails miserably. In such quantum and degenerate limit, $\eta \gg k_B T$, there is a possibility of quantum flux due to orbital levels transitions which reveals the traversing of chemical potential [21, 22, 38, 44, 49]. Here, the variation of chemical potential is responsible for charge density flux which leads to diffusion transport [29, 36]. This can be observed in our analysis in section of results and discussions in B and D (also, see FIG. 2 and 6). In this condition, the diode performance is so high at which the diode ideality factor takes a large deviation from unit, $g \gg 1$. Here, the enhancement of carrier current dominantly depends on the electronic term, i.e., chemical potential (or Fermi energy), rather than that of temperature (see Eq. 10 and Eq. 25). Performance wise diode functionality can be explained by our extended Einstein formalism at different physical conditions. For instance, at pure quantum limit, the ideality factor for any diode equations is a function of only one single parameter, chemical potential (see Eq. 11 and Eq. 27). Generally, the chemical potential is a direct evidence of presence of carrier density [5, 49]. Using our extended model, the diode equation can be incorporated with current density-voltage (*J-V*) analysis. Thus, now the diode equation of Schottky junction for degenerate materials becomes,

$$J = J_0 \left\{ \exp\left[ \frac{e(V - IR_S)}{\eta \left[ \frac{\left[1 + \exp\left(-\frac{2\eta^2}{\sigma^2 k_B^2 T^2}\right)\right]}{\left[1 + \exp\left(-\frac{2\eta^2}{\sigma^2 k_B^2 T^2}\right)\right] - \left(\frac{2\eta}{\sigma k_B T}\right)^2 \exp\left(-\frac{2\eta^2}{\sigma^2 k_B^2 T^2}\right)} \right]} \right] \right\}, \quad (33)$$

where, $J_o$, I, $R_S$, $\eta$, T and $\sigma$ are the saturation current density, the current, the series resistance, the chemical potential, the temperature and the Gaussian disorder width, respectively. Here, the disorder width is directly related to the existence of degenerate levels due to applied electric field (Stark effect) and magnetic field (Zeeman effect), see Refs. 21, 22 and 44. The chemical potential also can be modified by external field strength. In the case of $\sigma \rightarrow 0$, the Gaussian disordered function is equivalent to that of Dirac-delta function and now the diode equation (Eq. 33) is reduced as,

$$J = J_0 \left\{ \exp\left[ \frac{e(V - IR_S)}{\eta} \right] \right\} \quad (34)$$

In principle, the chemical potential depends on carrier dopants, bias, interface and contact effects and temperature. On the basis of entropy effect on D/µ ratio (see Eq. 25 and Eq. 26), the above diode Eq. 33 and Eq. 34 can be modified. We preserve the original Shockley diode equation for nondegenerate materials at high temperature regime (see Eq. 6, 7 and Eq. 12).

### A. Experimental verification

The earlier experimental study on chemical potential and quantum Hall Ferromagnetism in bilayer graphene (performed by Lee *et al.*) clearly indicate the relationship between chemical potential and quantum feature in the bilayer graphene system [44]. In this study, orbital level transitions were estimated using chemical potential, which is the analogue of Landau levels shifting. Also, the authors have calculated the carrier effective mass using electronic compressibility method. Due to many-body interactions, the existence of non-parabolic energy-momentum dispersion in this bilayer graphene strongly suggests the charge density (or electronic compressibility) dependent effective mass [44]. In this paper, the effective mass was calculated by electronic compressibility approach, $m^* = \frac{\pi\hbar^2}{2}\left(\frac{dn}{d\eta}\right)$. Here, the measurements were performed at T = 1.5 K. In this degenerate limit, we can formulate effective mass equation using our derivation (see Eq. 8 in section II). Without considering the disorder width (σ) in Eq.8, we can get the above effective mass equation. Using experimental data for the graphene (Sample # 1, Sample #2 and Sample # 3) [44], we have calculated the density dependent effective mass and D/μ relation using our extended Einstein model, which is consistent with the experimental findings (see Figs. 2 and 8). Thanks to Prof. E. Tutuc for sharing the experimental data.

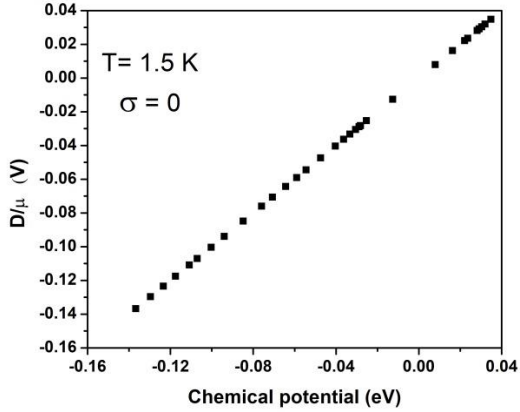

FIG. 8. Diffusion-mobility ratio with respect to the chemical potential for zero Gaussian disorder width at T = 1.5 K. The fitted data are related to sample #1: graphene (see Ref. 44).

In such a way, we have verified our model for layered $Bi_2Se_3$ materials also [43]. Using band diagram of $Bi_2Se_3$ (see FIGS. 2 (c) and 4(c) in Ref. 43), we can extract the carrier density versus chemical potential relationship. This experiment was performed at the temperature T = 1.8 K. According to our model, the calculated effective mass is 0.03 $m_0$.

### B. Band structure dependent D/μ

The enhancement of D/μ with respect to the graphene band structure was numerically calculated by Ancona (see FIG. 4 in Ref. 7). The calculated electron density and its relevant D/μ values clearly emphasis the importance of our extended Einstein model. It has been observed that there is a significant enhancement in D/μ relation even at room temperature, which strongly depends on the carrier density (or chemical potential). Using our model, the calculated D/μ enhancement at different temperature for zero disorder width is plotted which is shown in FIG. 9.

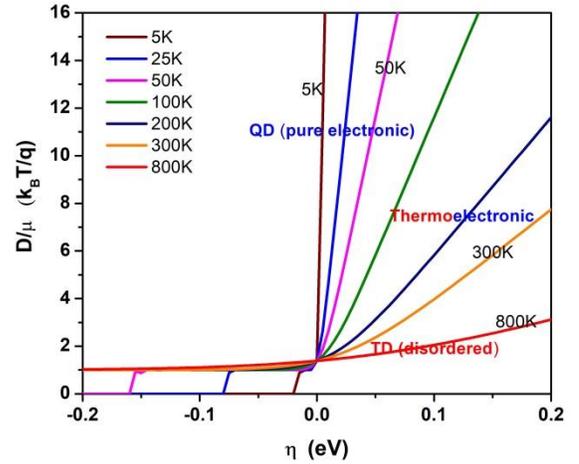

FIG. 9. The enhancement of D/μ with respect to the chemical potential for different temperature values at zero disorder width.

We have preserved the original Einstein relation from our extended model for wide range of negative chemical potential (localized charge transport regime) at room temperature and the above. Importantly, the validity of Einstein relation is observed for entire chemical potential range (including positive values of chemical potential, which is the delocalized regime) at very high temperature range, e.g., T = 800 K and the above, see Figs. 1 and 9. Here, the chemical potential is mainly responsible for charge density. We have verified our model basis D/μ transport with the earlier band structure versus D/μ study for graphene (see FIG. 4 in Ref. 7). Our model basis D/μ enhancement at room temperature follows the similar trend as in Ref. 7. This is the one of the direct evidences for our extended D/μ transport model.

## V. SUMMARY AND CONCLUSIONS

Using our extended Einstein model, we have studied the diffusion-mobility transport at different physical domain for generic and degenerate Schrödinger systems. For certain combinations of temperature and chemical potential, we preserve the original Einstein equation from our model. At high chemical potential (delocalized region), there is D/μ deviation from the original Einstein value of $k_B T/e$, which implies the D/μ enhancement. In strongly localized limit (η = -∞, leads to zero density), the absence of diffusion is observed which is the insulator region. For degenerate materials, the chemical potential is a deterministic parameter in D/μ transport. The electron-hole symmetrical transport is observed for degenerate ideal quantum materials. The existing degeneracy levels, due to applied electric or magnetic fields, significantly modifies the DOS, accordingly the D/μ is quite enhanced. The shape and width of DOS and its consequences on D/μ enhancement in such degenerate limits is quantified from Gaussian variance and chemical potential. This work clearly emphasis that the presence of many-body entropy strongly disturbs the diffusion and hence it completely shuts the diffusion-mobility transport, even in degenerate cases for localized domain (for negative chemical potential region). It is interestingly observed that the carrier dynamics starts with one jerk based dynamics (kind of oscillation) in entropy modulated Fermi systems and thereafter it takes continues smooth D/μ enhancement with respect to the chemical potential. This sudden down-up peak at finite chemical potential (or oscillation with small window) suggests that the entropy inducts dipole states through the temperature mediated polaronic effects. In practical devices, the possibility of defect states and disorder shuts the transport route during diffusion process (zero diffusion), in which the increased entropy term maximize the localization strength. We noted the thermal fluctuation in entropy modulated D/μ transport for generic Fermi systems; on the other hand, the quantum fluctuation is appeared in this transport for entropy modulated degenerate Fermi systems. We importantly infer two corollaries; 1) the diffusion-mobility is thermally activated in localized transport region (e.g., organic photovoltaic solar cells), 2) diffusion is limited by thermal disorder in delocalized region, due to temperature mediated scattering. These findings are in agreement with the earlier observations [36, 39, 41]. Also, we have verified our formalism with the earlier experimental and theoretical results of different quantum degenerate materials like, graphene, $Bi_2Se_3$, and a few. Based on the analysis, we have shown the validity and importance of our extended Einstein model for the advancement in nanoelectronic devices. Based on our extended Einstein equation, we have explicitly revisited the well-known Shockley diode equation as $J = J_0 \left[ \exp\left(\frac{eV}{\eta} - 1\right) \right]$ for ideal degenerate Schrödinger type quantum materials. This ideal quantum diode equation directly drags current density enhancement, empirically fitted by ideality factor, which was reported in earlier studies (see FIG. 4 in Ref. 7) [7, 8, 15, 50].

## ACKNOWLEDGEMENTS
The author is grateful to Prof. Emanuel Tutuc and Dr. Kayoung Lee for sharing their experimental data of "Science **345**, 58 (2014)" to validate our model. Also, the author is thankful to Prof. Swapan K. Pati, Dr. K. R. Vijayaraghavan, Dr. V. Govindaraj, and Dr. Bradraj Pandey for their useful discussions on this topic. Also, the author wishes to acknowledge Dr. K. Rajkumar and Dr. Shubhajit Das for their suggestions in manuscript preparation.

## APPENDIX A: Carrier density, electronic compressibility and diffusion-mobility (D/μ) relation for two-dimensional (2D) Schrödinger materials

The number of charge carrier for degenerate 2D materials in nonrelativistic limit can be derived using Fermi-Dirac (FD) distribution and Gaussian density of states

$$N_{2D} = \int_{-\infty}^{+\infty} f(E) g(E)_{2D} dE \quad (A1)$$

where, $f(E)$ is the Fermi-Dirac distribution function, $f(E) = \frac{1}{1 + exp\left(\frac{E-\eta}{k_B T}\right)}$ and $g(E)$ is the Gaussian density of states, $g(E) = \frac{N_g}{\sqrt{2\pi}\sigma} exp\left(-\frac{\varepsilon^2}{2\sigma^2}\right)$. Here $\eta$ is the chemical potential, $N_g$ is the total density of states, $\varepsilon$ is the normalized energy $= \frac{E}{k_B T}$, and $\sigma$ is the normalized Gaussian variance $\sigma = \frac{\sigma_0}{k_B T}$ (or width).

For 2D materials, the total density of states is generally defined as, $N_{g,2D} = \frac{\partial N_{st,2D}}{\partial E} = \frac{mL^2}{\pi \hbar^2}$. Here, $N_{st,2D}$ is the total number of states, $m$ is the effective mass of the carrier, $L$ is the length of the box (or system) and $\hbar$ is the reduced Planck constant.

According to Eq. (1), the number of charge carrier can be expressed as,

$$N_{2D} = \frac{2mL^2}{\pi\sqrt{2\pi}\hbar^2} \frac{1}{\sigma} \int_0^{+\infty} \frac{1}{1 + exp\left(\frac{E-\eta}{k_B T}\right)} exp\left(-\frac{\varepsilon^2}{2\sigma^2}\right) dE \quad (A2)$$

Here, $\beta = \frac{1}{k_B T}$ (Lagrange multiplier).

*K. Navamani*

We use the method of $\int u.dv = uv - \int v.du$ and solve the Eq. (A2). Let us consider, $u = exp\left(-\frac{\varepsilon^2}{2\sigma^2}\right)$ and $dv = \frac{1}{1+exp(\beta(E-\eta))}dE$. Thus,

$$v = \int dv = \int_0^{+\infty} \frac{1}{1+exp(\beta(E-\eta))}dE = \int_0^{+\infty} \frac{exp(-\beta(E-\eta))}{1+exp(-\beta(E-\eta))}dE \tag{A3}$$

Let us take $x = exp(-\beta(E-\eta))$. Here $x$ and $E$ are variables. Take the logarithm both sides and differentiate it, then we get $log\, x = -\beta(E-\eta) \Rightarrow \frac{dx}{x} = -\beta dE$. Now the lower and upper limits of integration are $exp(\beta\eta)$ and 0, respectively.

$$v = \int_{exp(\beta\eta)}^{0} \frac{x}{1+x}\left(-\frac{dx}{\beta x}\right) = -\frac{1}{\beta}\int_{exp(\beta\eta)}^{0} \frac{1}{1+x}dx = -\frac{1}{\beta}[ln(1+x)]_{exp(\beta\eta)}^{0} \tag{A4}$$

$$v = \frac{1}{\beta}ln[1 + exp(\beta\eta)] \tag{A5}$$

In such a way,

$$du = -\left(\frac{\varepsilon}{\sigma^2}\right)exp\left(-\frac{\varepsilon^2}{2\sigma^2}\right)d\varepsilon \tag{A6}$$

$$\int v.du = -\int_0^{+\infty} \frac{1}{\beta}ln[1+exp(\beta\eta)]\left[\left(\frac{\varepsilon}{\sigma^2}\right)exp\left(-\frac{\varepsilon^2}{2\sigma^2}\right)\right]d\varepsilon = -\frac{1}{\beta}ln[1+exp(\beta\eta)]\int_0^{+\infty}\left(\frac{\varepsilon}{\sigma^2}\right)exp\left(-\frac{\varepsilon^2}{2\sigma^2}\right)d\varepsilon \tag{A7}$$

Let us assume $y = exp\left(-\frac{\varepsilon^2}{2\sigma^2}\right)$, $ln\, y = -\frac{\varepsilon^2}{2\sigma^2} \Rightarrow \frac{dy}{y} = -\frac{\varepsilon.d\varepsilon}{\sigma^2}$
In this case, the lower and upper limits of integration are 1 and 0, respectively. Now Eq. (A7) can be written as,

$$\int v.du = -\frac{1}{\beta}ln[1+exp(\beta\eta)]\int_1^0 y\left(-\frac{dy}{y}\right) = -\frac{1}{\beta}ln[1+exp(\beta\eta)]\,[-y]_1^0$$

$$\int v.du = -\frac{1}{\beta}ln[1+exp(\beta\eta)] \tag{A8}$$

$$\int u.dv = uv - \int v.du$$

Thus,

$$\int_0^{+\infty} \frac{1}{1+exp(\beta(E-\eta))}exp\left(-\frac{\varepsilon^2}{2\sigma^2}\right)dE = \frac{1}{\beta}exp\left(-\frac{\varepsilon^2}{2\sigma^2}\right)ln[1+exp(\beta\eta)] + \frac{1}{\beta}ln[1+exp(\beta\eta)] \tag{A9}$$

The simplified form of Eq. (A9) can be written as,

$$\int_0^{+\infty} \frac{1}{1+exp(\beta(E-\eta))}exp\left(-\frac{\varepsilon^2}{2\sigma^2}\right)dE = \frac{1}{\beta}ln(1+exp(\beta\eta))\left[1+exp\left(-\frac{\varepsilon^2}{2\sigma^2}\right)\right] \tag{A10}$$

Inserting Eq. (A10) in to Eq. (A2) and we can be expressed the carrier density of 2D degenerate systems as,

$$n_{2D,GDOS} = \frac{N_{2D}}{L^2} = \frac{2mk_BT}{\pi\sqrt{2\pi}\hbar^2}\frac{1}{\sigma}ln\left(1+exp\left(\frac{\eta}{k_BT}\right)\right)\left[1+exp\left(-\frac{\varepsilon^2}{2\sigma^2}\right)\right] \tag{A11}$$

This formula describes the total distribution of charge carrier density in the Gaussian density of states. For nondegenerate cases (Maxwellian form) Eq. (A11) is reduced as,

$$n_{2D} = \frac{N_{2D}}{L^2} = \frac{2mk_BT}{\pi\sqrt{2\pi}\hbar^2}\frac{1}{\sigma}exp\left(\frac{\eta}{k_BT}\right)\left[1 + exp\left(-\frac{\varepsilon^2}{2\sigma^2}\right)\right] \tag{A12}$$

Without using the Gaussian function, $\frac{1}{\sqrt{2\pi}\sigma}exp\left(-\frac{\varepsilon^2}{2\sigma^2}\right)$, one can generally estimate the number of carrier by following manner,

$$N_{2D} = \int_{-\infty}^{+\infty} DOS_{2D}\, f(E)dE = \frac{2m}{\pi\hbar^2}\int_0^{+\infty}\frac{1}{1+exp\left(\frac{E-\eta}{k_BT}\right)}dE \tag{A13}$$

In Eq. (A13), $DOS_{2D} = N_{g,2D} = \frac{\partial N_{st,2D}}{\partial E} = \frac{mL^2}{\pi\hbar^2}$. Now the value of carrier density becomes,

$$n_{2D} = \frac{2mk_BT}{\pi\hbar^2}ln\left(1 + exp\left(\frac{\eta}{k_BT}\right)\right) \tag{A14}$$

Normally, the relation between the carrier density and wave vector for 2D system is defined as,

$$n_{2D} = \frac{1}{2\pi}k_F^{\,2} \tag{A15}$$

By comparing Eq. (A14) and Eqn. (A15),

$$\frac{2mk_BT}{\pi\hbar^2}ln\left(1 + exp\left(\frac{\eta}{k_BT}\right)\right) = \frac{1}{2\pi}k_F^{\,2} \tag{A16}$$

Finally, we get the particle's kinetic energy in terms of chemical potential and of temperature, and it can be described from Eq. (A16) as,

$$E = \frac{p^2}{2m} = 2k_BT\, ln\left(1 + exp\left(\frac{\eta}{k_BT}\right)\right) \tag{A17}$$

Thus, the normalized energy is

$$\varepsilon = \frac{E}{k_BT} = 2\, ln\left(1 + exp\left(\frac{\eta}{k_BT}\right)\right) \tag{A18}$$

Using Eq. (A18), the Gaussian disordered charge density Eq. (A11) can be modified as,

$$n_{2D} = \frac{2mk_BT}{\pi\hbar^2\sqrt{2\pi}}\frac{1}{\sigma}ln\left(1 + exp\left(\frac{\eta}{k_BT}\right)\right)\left[1 + exp\left(-\frac{2\left(ln\left(1+exp\left(\frac{\eta}{k_BT}\right)\right)\right)^2}{\sigma^2}\right)\right] \tag{A19}$$

Now, the electronic compressibility can be explicitly described as,

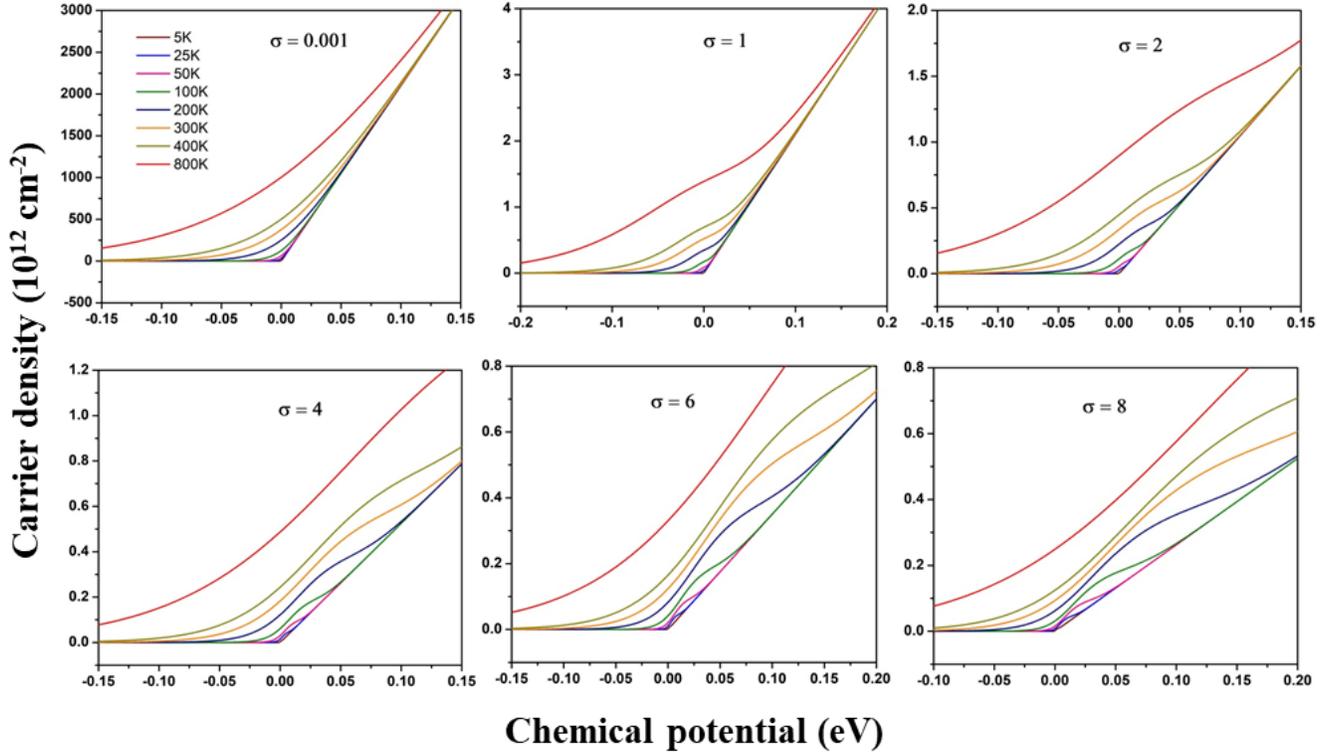

FIG. A1. Carrier density as a function of chemical potential at different temperature values for different Gaussian variances. The localized carriers are activated by temperature.

$$\frac{\partial n_{2D}}{\partial \eta} = \frac{2m}{\pi\hbar^2\sqrt{2\pi}}\frac{1}{\sigma}\frac{exp\left(\frac{\eta}{k_BT}\right)}{\left(1+exp\left(\frac{\eta}{k_BT}\right)\right)}\left\{\left[1+exp\left(-\frac{2\left(ln\left(1+exp\left(\frac{\eta}{k_BT}\right)\right)\right)^2}{\sigma^2}\right)\right]-\frac{4\left(ln\left(1+exp\left(\frac{\eta}{k_BT}\right)\right)\right)^2}{\sigma^2}exp\left[-\frac{2\left(ln\left(1+exp\left(\frac{\eta}{k_BT}\right)\right)\right)^2}{\sigma^2}\right]\right\} \quad \text{(A20)}$$

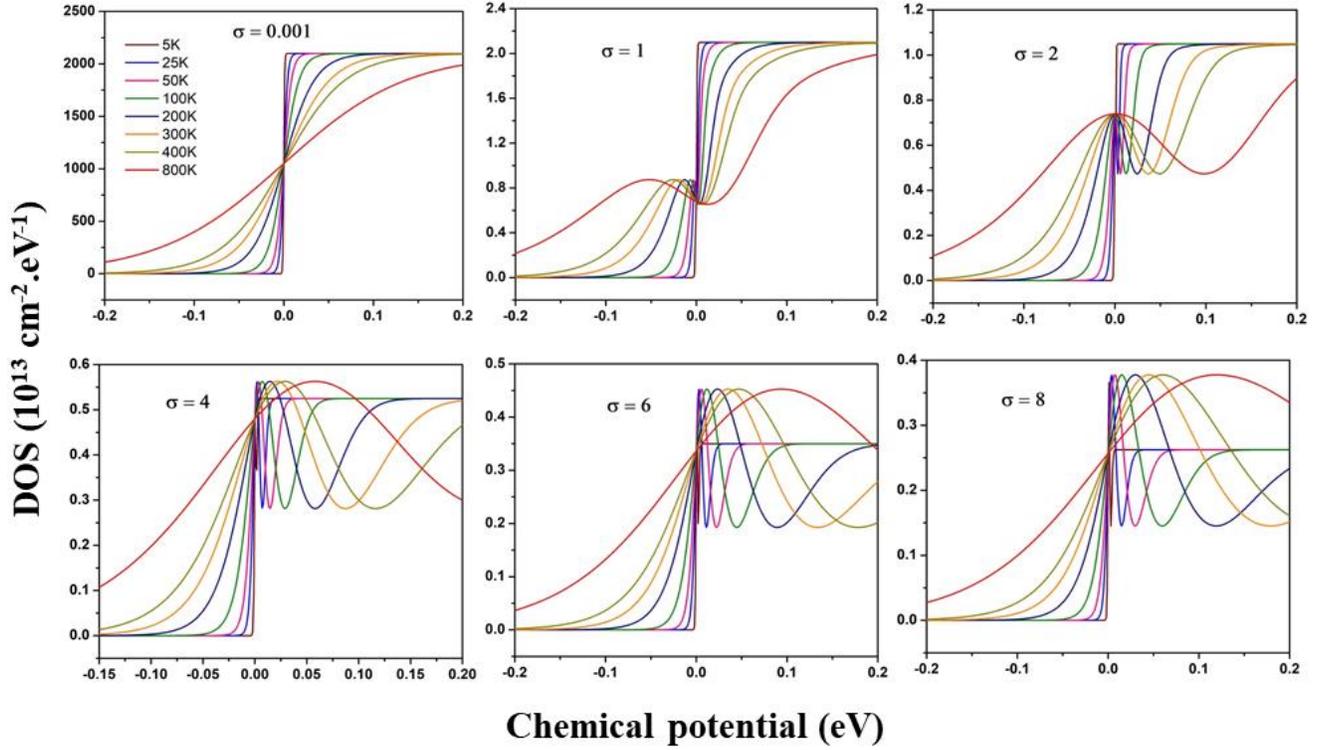

FIG.A2. DOS (or electronic compressibility) and its shape are depending upon chemical potential (or charge transport energy), temperature, and Gaussian variance. In localized region ($\eta < 0$), the expected DOS decreases with the Gaussian variance, which can be controlled by applied electric (Stark effect) and magnetic field (Zeeman effect).

General Einstein D/μ relation is defined as [5, 37],

$$\left(\frac{D}{\mu}\right) = \frac{n}{e(\partial n/\partial \eta)} \tag{A21}$$

By inserting Eq. (A19) and (A20) in to Eq. (A21), we get the generalized expression of D/μ for 2D-Schrödinger materials and it can be explicitly written as,

$$\left(\frac{D}{\mu}\right)_{2D} = \frac{k_B T}{e}\left[\frac{\left(1+exp\left(\frac{\eta}{k_B T}\right)\right)ln\left(1+exp\left(\frac{\eta}{k_B T}\right)\right)}{exp\left(\frac{\eta}{k_B T}\right)}\right]\left\{\frac{\left[1+exp\left(-\frac{2\left(ln\left(1+exp\left(\frac{\eta}{k_B T}\right)\right)\right)^2}{\sigma^2}\right)\right]}{\left[1+exp\left(-\frac{2\left(ln\left(1+exp\left(\frac{\eta}{k_B T}\right)\right)\right)^2}{\sigma^2}\right)\right]-\frac{4\left(ln\left(1+exp\left(\frac{\eta}{k_B T}\right)\right)\right)^2}{\sigma^2}exp\left[-\frac{2\left(ln\left(1+exp\left(\frac{\eta}{k_B T}\right)\right)\right)^2}{\sigma^2}\right]}\right\} \tag{A22}$$

To this connection, the survival time of charge carrier can be estimated via the uncertainty relation as, $t = \frac{\hbar\mu}{2eD}$. Here, D/μ factor is equivalent to that of carrier potential.

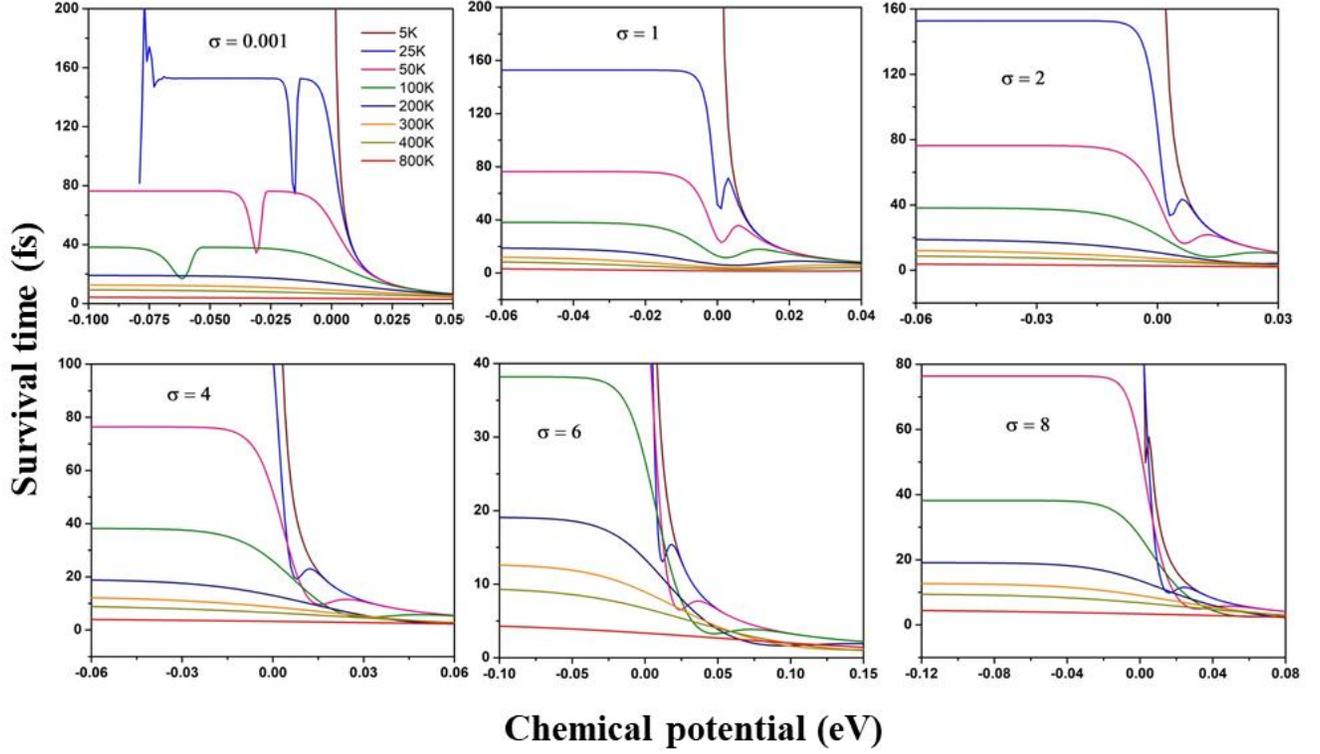

FIG. A3. Survival time directly gives the validity region of Einstein transport equation, $\frac{D}{\mu} = \frac{k_B T}{e}$, and invalidity region of Einstein transport equation, $\frac{D}{\mu} > \frac{k_B T}{e}$.

In zero dispersion (too weak Gaussian disorder), $\sigma \rightarrow 0$, the above general form of D/μ for 2D-Schrödinger materials becomes,

$$\left(\frac{D}{\mu}\right)_{2D} = \frac{k_B T}{e} \left[ \frac{\left(1+\exp\left(\frac{\eta}{k_B T}\right)\right) \ln\left(1+\exp\left(\frac{\eta}{k_B T}\right)\right)}{\exp\left(\frac{\eta}{k_B T}\right)} \right] \tag{A23}$$

For degenerate quantum cases ($\eta \gg k_B T$), the Eq. (A19), (A20) and Eq. (A22) are reduced as,

$$n_{2D} = \frac{2m\eta}{\pi \hbar^2 \sqrt{2\pi}} \frac{1}{\sigma} \left[ 1 + \exp\left(-\frac{2\eta^2}{\sigma^2 k_B^2 T^2}\right) \right] \tag{A24}$$

$$\frac{\partial n_{2D}}{\partial \eta} = \frac{2m}{\pi \hbar^2 \sqrt{2\pi}} \frac{1}{\sigma} \left\{ \left[ 1 + \exp\left(-\frac{2\eta^2}{\sigma^2 k_B^2 T^2}\right) \right] - \frac{4\eta^2}{\sigma^2 k_B^2 T^2} \exp\left[-\frac{2\eta^2}{\sigma^2 k_B^2 T^2}\right] \right\} \tag{A25}$$

$$\frac{D}{\mu} = \frac{\eta}{e} \left\{ \frac{\left[1+\exp\left(-\frac{2\eta^2}{\sigma^2 k_B^2 T^2}\right)\right]}{\left[1+\exp\left(-\frac{2\eta^2}{\sigma^2 k_B^2 T^2}\right)\right] - \left(\frac{2\eta}{\sigma k_B T}\right)^2 \exp\left(-\frac{2\eta^2}{\sigma^2 k_B^2 T^2}\right)} \right\} \tag{A26}$$

For very weak disorder ($\sigma \rightarrow 0$), or zero disorder ($\sigma = 0$) in degenerate materials, the D/μ equation becomes,

*K. Navamani*

$$\left(\frac{D}{\mu}\right)_{2D} = \frac{\eta}{e} \tag{A27}$$

The Eq. (A27) is the fundamental transport equation for quantum materials and it works very well at very low temperature regime. The above quantum diffusion-mobility relation purely depends on chemical potential (or Fermi energy at zero temperature).

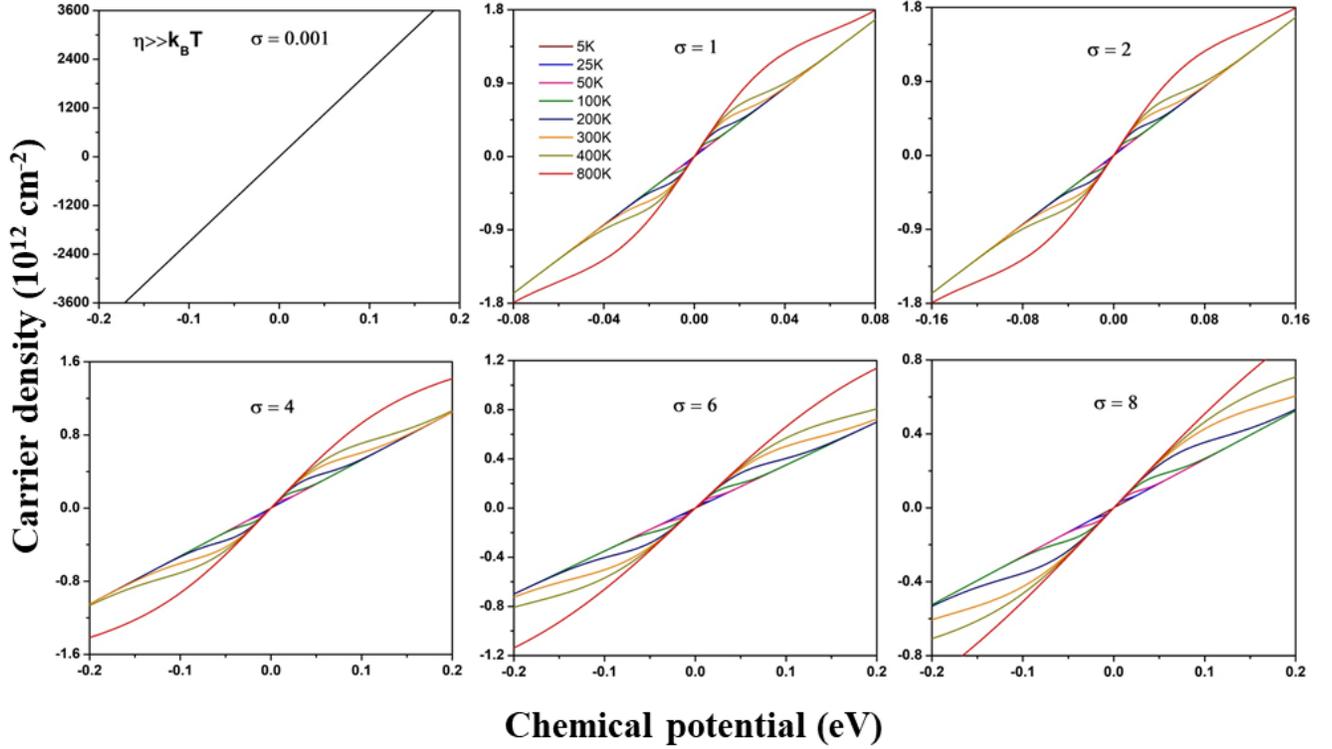

FIG. A4. Carrier density profile (for degenerate 2D systems) as a function of chemical potential at different temperature values for different Gaussian variances. Electron and hole densities are estimated from positive region and from negative region of chemical potential values, respectively. The patterns follow the inversion symmetry rule.

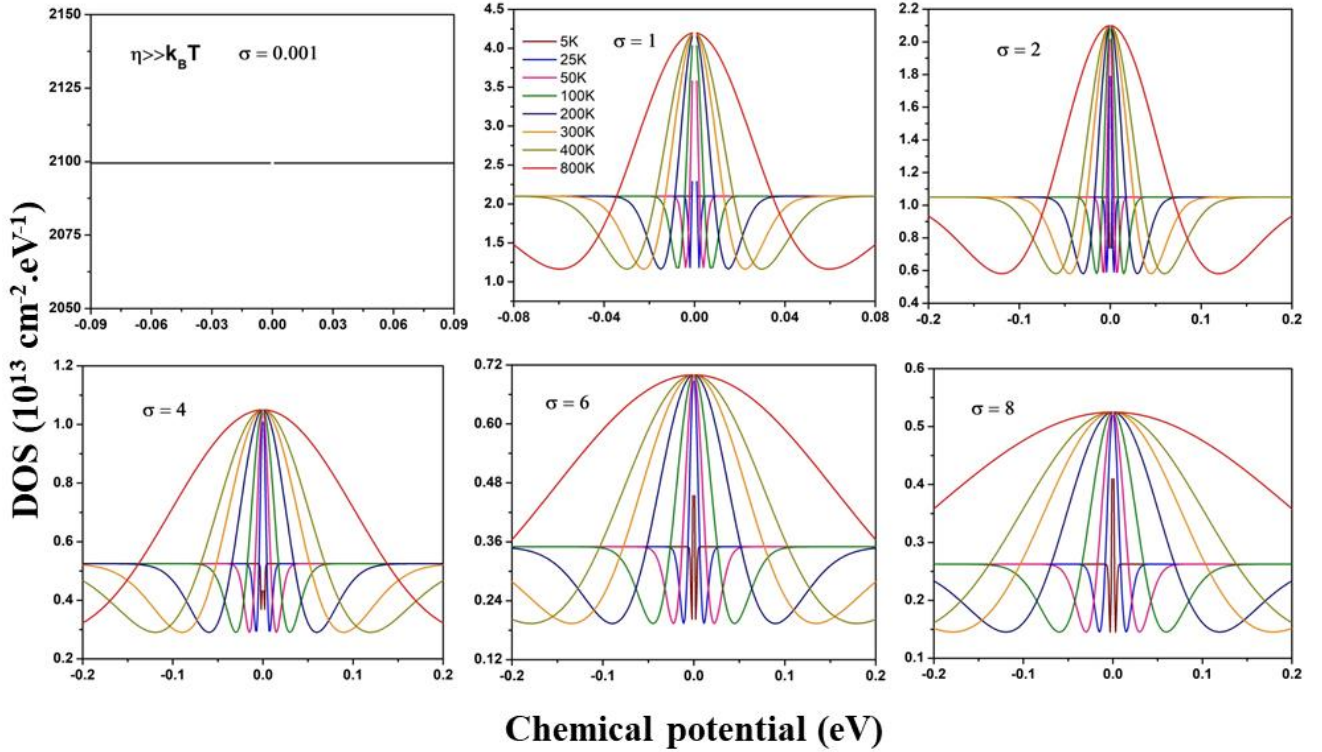

FIG. A5. DOS and its distributions in the wide chemical potential range at different Gaussian variance values. The existence of mirror symmetry shows the possibility of electron-hole symmetrical in degenerate ideal 2D Schrödinger systems.

## APPENDIX B
**Entropy effect on carrier density and on diffusion coefficient**

According to density flux model, the disorder effect on carrier density in 3D materials can be described as [42],

$$n_{3D,S} = n_{3D,0}\, exp\left(-\frac{3S}{5k_B}\right) \quad (B1)$$

Total energy of bulk materials (3D) as [51],

$$E_{3D} = \frac{\hbar^2(3\pi^2 N e)^{5/3}}{10\pi^2 m} V^{-2/3} = \frac{\hbar^2}{10\pi^2 m} k^5 V \quad (B2)$$

where, $N$ is the number of carrier, $V=L^3$ is the volume and $k$ is the wave vector ($k = 3\pi^2 n_{3D})^{1/3}$.

For 2D systems, the wave vector is $k = (2\pi n_{2D})^{1/2}$. Hence, the total energy for the 2D systems becomes,

$$E_{2D} = \frac{2\sqrt{2\pi}\hbar^2(n_{2D}A)^{5/2}}{5m} A^{-1} = \frac{2\sqrt{2\pi}\hbar^2(N_{2D})^{5/2}}{5m} A^{-1} \quad (B3)$$

Here, $A$ is the area, $A=L^2$, and $N_{2D}$ is the number of carriers.

By the external work done on the system, the change of energy at small interval of time can be defined as,

$$\frac{dE_{2D}}{dt} = -\frac{2\sqrt{2\pi}\hbar^2(N_{2D})^{5/2}}{5m} A^{-2} \frac{dA}{dt} = -\frac{2\sqrt{2\pi}\hbar^2(N_{2D})^{1/2}}{5m}\left(\frac{N_{2D}}{A}\right)^2 \frac{dA}{dt} \quad (B4)$$

Here, we assume that the number of carriers is conserved. In the high temperature, the lattice distortion and the nuclear dynamics significantly alter the carrier distribution area which leads to non-equilibrium. In such that the dynamic disorder due to lattice dynamics limits the electronic transport via diffusion [39, 52]. Thus, the carrier flux rate (or redistribution speed) must be associated with the spatial distribution (or area) changes, which is responsible for deformation potential.

The simplified form of Eq. (B4) as,

$$\frac{dE_{2D}}{dt} = -B n_{2D}^2 \frac{dA}{dt} \quad (B5)$$

where, $B$ is the constant, $B = \frac{2\sqrt{2\pi}\hbar^2(N_{2D})^{1/2}}{5m}$. The negative sign in Eq. (B5) represents the energy flux into the system due to external work done, i.e., energy flow from environment to system (e.g., temperature, electric field, etc.). In our case, we assume the constant perturbation with time; hence the change in spatial area with respect to the time is constant. Here, the external interactions have uniformly disturbed all N-particles in a system at equal interval of time, which leads to thermal averaging effect at each time interval. Hence, $\frac{dA}{dt}$ from Eq. (B5) is the constant value.

The generalized form of "disorder (in terms of entropy) controlled shuttling energy rate equation" can be expressed as [11],

$$\left(\frac{\partial E}{\partial t}\right)_S = \left(\frac{\partial E}{\partial t}\right)_0 exp\left(-\frac{S}{k_B}\right) \quad (B6)$$

By comparing Eq. (B5) and Eq. (B6), the disorder dependent density can be achieved for 2D materials as,

$$n_{2D,S} = n_{2D,0}\, exp\left(-\frac{S}{2k_B}\right) \quad (B7)$$

The above equation relates the entropy limited carrier density for electronic transport, which quantitatively affects the conductivity.

Here the shuttling energy determines the charge carrier flux in the systems. According to the Poisson's equation, the second order derivative of potential (due to applied voltage) can be written as,

$$\frac{\partial^2 V}{\partial X^2} = -\frac{\rho_{3D}}{\varepsilon} \quad (B8)$$

where, $\rho_{3D}(=n_{3D}e)$ and $\varepsilon$ are charge density and electric permittivity of the medium, respectively.

According to earlier descriptions [11, 53], the density flux leads to potential difference which is related with the diffusion mechanism, and hence it can be written as,

$$\frac{\partial V}{\partial t} = D\frac{d^2 V}{dX^2} \quad (B9)$$

Inserting the Eq. (B8) in to Eq. (B9), and one can get,

$$\frac{\partial V}{\partial t} = -\frac{D\rho_{3D}}{\varepsilon} = -\left(\frac{e}{\varepsilon}\right)Dn_{3D} \quad (B10)$$

The other form of Eq. (B10) can be written as,

$$\frac{\partial E}{\partial t} = \frac{e^2 D n_{3D}}{\varepsilon} = \frac{e^2 D k^3}{3\pi^2 \varepsilon} \quad (B11)$$

The shuttling energy rate for 2D system can be defined as,

$$\frac{\partial E}{\partial t} = \frac{e^2 D(2\pi n_{2D})^{3/2}}{3\pi^2 \varepsilon} = \frac{e^2 (2\pi)^{3/2}}{3\pi^2 \varepsilon} D(n_{2D})^{3/2} \quad (B12)$$

By comparing Eq. (B6) and Eq. (B12), we can be derived the following relation as,

$$D_{2D,S}(n_{2D,S})^{3/2} = D_{2D,0}(n_{2D,0})^{3/2} exp\left(-\frac{S}{k_B}\right) \quad (B13)$$

Inserting the Eq. (B7) in to Eq. (B13), and finally we obtain the equation of disorder dependent diffusion coefficient in 2D materials and it can be expressed as,

$$D_{2D,S} = D_{2D,0}\, exp\left(-\frac{S}{4k_B}\right) \quad (B14)$$

Here, $D_{2D,0}$ is the diffusion coefficient at zero disorder ($S=0$). The Eq. (B14) is the entropy dependent diffusion equation which originally describes how the diffusion current is limited by thermal disorder. This description is in conceptually good agreement with the Troisi's studies [39, 40, 46].

## APPENDIX C
**Entropy derivation for Schrödinger materials**

For degenerate 2D materials, the entropy is generally defined from imperfect Fermion gas and it can be expressed as [35],

$$S = \frac{\pi^3}{3}\frac{k_B^2 T}{\sqrt{2\pi n_{2D}}}\left(\frac{\partial E_k}{\partial k}\right)^{-1} \quad (C1)$$

$$S = \frac{\pi^3}{3}\frac{k_B^2 T}{k}\frac{m}{\hbar^2 k} = \frac{\pi^3}{3}k_B^2 T\frac{m}{p^2} = \frac{\pi^3}{6}k_B^2 T\frac{1}{E_K} \quad (C2)$$

Substituting Eq. (A17) in to Eq. (C2), entropy can be expressed as,

$$S = \frac{\pi^3}{12}k_B \frac{1}{ln\left(1+exp\left(\frac{\eta}{k_B T}\right)\right)} \quad (C3)$$

In quantum limit of $\eta \gg k_B T$, the entropy formula will be simplified as,

$$S = \frac{\pi^3}{12}k_B \frac{k_B T}{\eta} \quad (C4)$$

In similar way, the entropy for nondegenerate cases (Maxwellian form) can be obtained as,

$$S = \frac{\pi^3}{12}k_B\, exp\left(-\frac{\eta}{k_B T}\right) \quad (C5)$$

It is to be noted that the entropy can be quantified by the combination of thermal energy and chemical potential. Generally, electronic part can be analyzed by chemical potential [38, 43, 44]. That is, based on the value of electronic and thermal components in a system, one can estimate the entropy of a particular system. Interestingly, we find that the entropy is a linear proportional with the thermal energy and a inversely linear proportional with the chemical potential for quantum materials. For nondegenerate materials, both the temperature and chemical potential are nonlinearly related to the entropy.

## APPENDIX D

**Entropy modulated Gaussian charge distribution model and its consequences on electronic compressibility and on D/μ relation (for Schrödinger materials)**

Inserting Eq. (A14) and (C3) in to Eq. (B7), we can be expressed the entropy-dependent carrier density for generalized 2D materials,

$$n_{2D,S} = \frac{2mk_BT}{\pi\hbar^2} \ln\left(1 + exp\left(\frac{\eta}{k_BT}\right)\right) exp\left[-\frac{\pi^3}{24 \ln\left(1+exp\left(\frac{\eta}{k_BT}\right)\right)}\right] \tag{D1}$$

Using carrier density-wave vector relationship $\left(k = \sqrt{2\pi n_{2D}}\right)$, one can derive the carrier energy from Eq. (D1) and it can be written as,

$$E_S = 2k_BT \ln\left(1 + exp\left(\frac{\eta}{k_BT}\right)\right) exp\left[-\frac{\pi^3}{24 \ln\left(1+exp\left(\frac{\eta}{k_BT}\right)\right)}\right] \tag{D2}$$

Thus, normalized energy becomes,

$$\varepsilon_S = 2 \ln\left(1 + exp\left(\frac{\eta}{k_BT}\right)\right) exp\left[-\frac{\pi^3}{24 \ln\left(1+exp\left(\frac{\eta}{k_BT}\right)\right)}\right] \tag{D3}$$

The above equation describes the limitation of charge transfer kinetics (charge transport energy) by thermal disorder, which can be explained by entropy (see APPENDIX C).

Using Eq. (A11) and (B7), the entropy modulated Gaussian carrier density equation can be written as,

$$n_{2D,S} = \frac{N_{2D,S}}{L^2} = \frac{2mk_BT}{\pi\sqrt{2\pi}\hbar^2}\frac{1}{\sigma} \ln\left(1 + exp\left(\frac{\eta}{k_BT}\right)\right) exp\left(-\frac{S}{2k_B}\right)\left[1 + exp\left(-\frac{\varepsilon_S^2}{2\sigma^2}\right)\right] \tag{D4}$$

Substituting Eq. (C3), (D3) in to Eq. (D4), the modified Gaussian carrier density equation can be explicitly defined as,

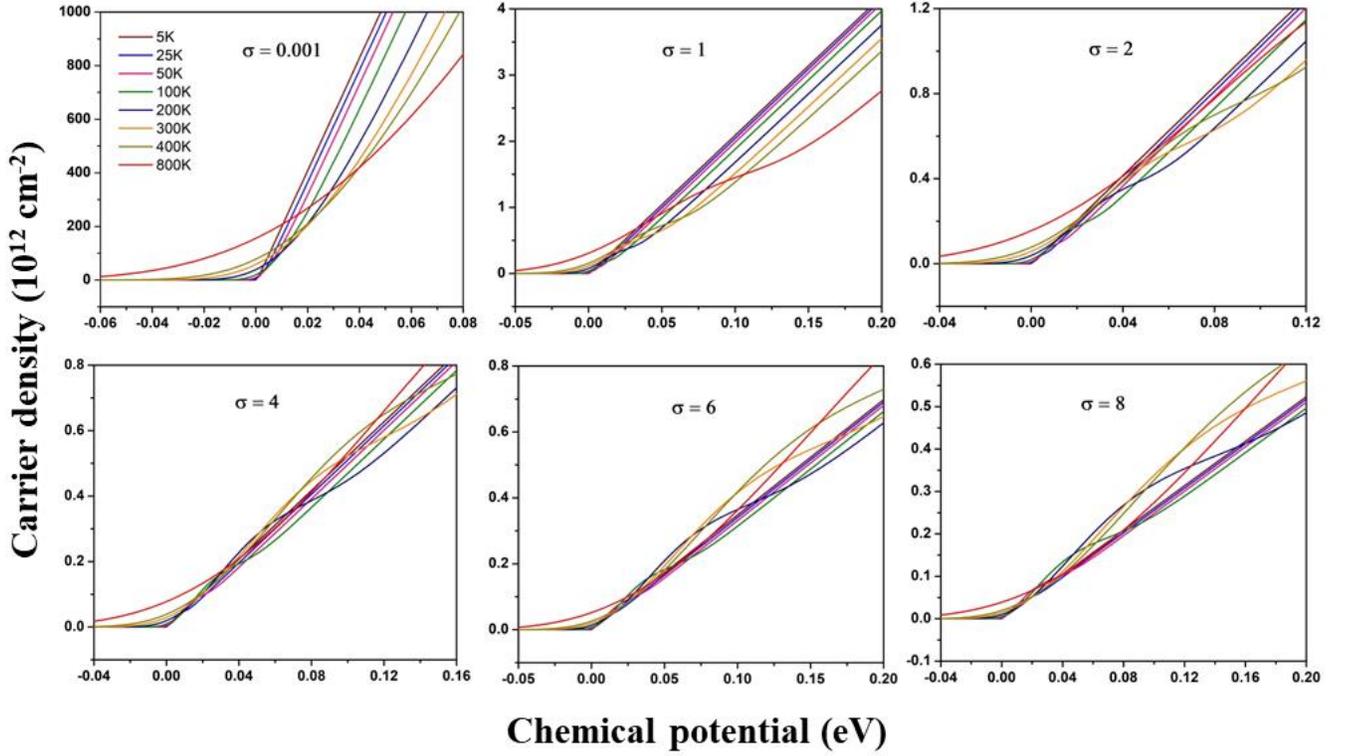

FIG. A6. Entropy modulated carrier density as a function of chemical potential at different temperature values for different Gaussian variances. The presence of entropy enhances the carrier localization in localization region and thus limits the contribution of carrier density value for electronic transport. In this domain (η < 0), the charge carrier is activated by temperature. In the delocalized domain (η > 0), the amount of carrier density strongly depends on cooperative behavior between the Gaussian width, the chemical potential and temperature.

$$n_{2D,S} = \frac{2mk_BT}{\pi\hbar^2\sqrt{2\pi}}\frac{1}{\sigma}ln\left(1+exp\left(\frac{\eta}{k_BT}\right)\right)exp\left[-\frac{\pi^3}{24\,ln\left(1+exp\left(\frac{\eta}{k_BT}\right)\right)}\right]\left\{1+exp\left[-\frac{2\left[ln\left(1+exp\left(\frac{\eta}{k_BT}\right)\right)\right]^2 exp\left(-\frac{\pi^3}{12\,ln\left(1+exp\left(\frac{\eta}{k_BT}\right)\right)}\right)}{\sigma^2}\right]\right\} \quad (D5)$$

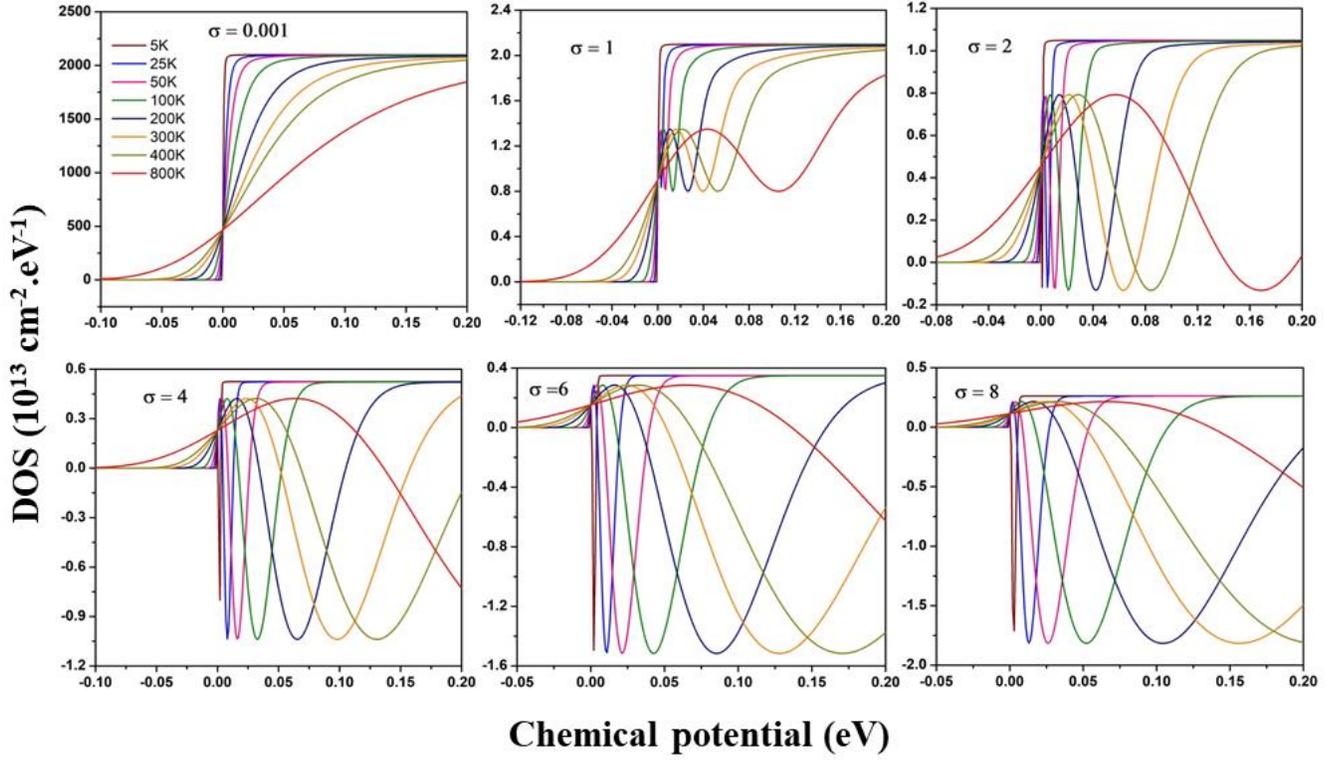

FIG. A7. Entropy modulated DOS (or electronic compressibility, see Eq. (D6)) and its shape are depending upon chemical potential (or charge transport energy), temperature, and Gaussian variance. The presence of entropy significantly suppresses the DOS in localized region ($\eta < 0$), which indicates the entropy assisted localization property and thus absence of diffusion is expected. On the other hand ($\eta > 0$), the entropy effect gives rise to the negative DOS values (or negative compressibility) for larger Gaussian width.

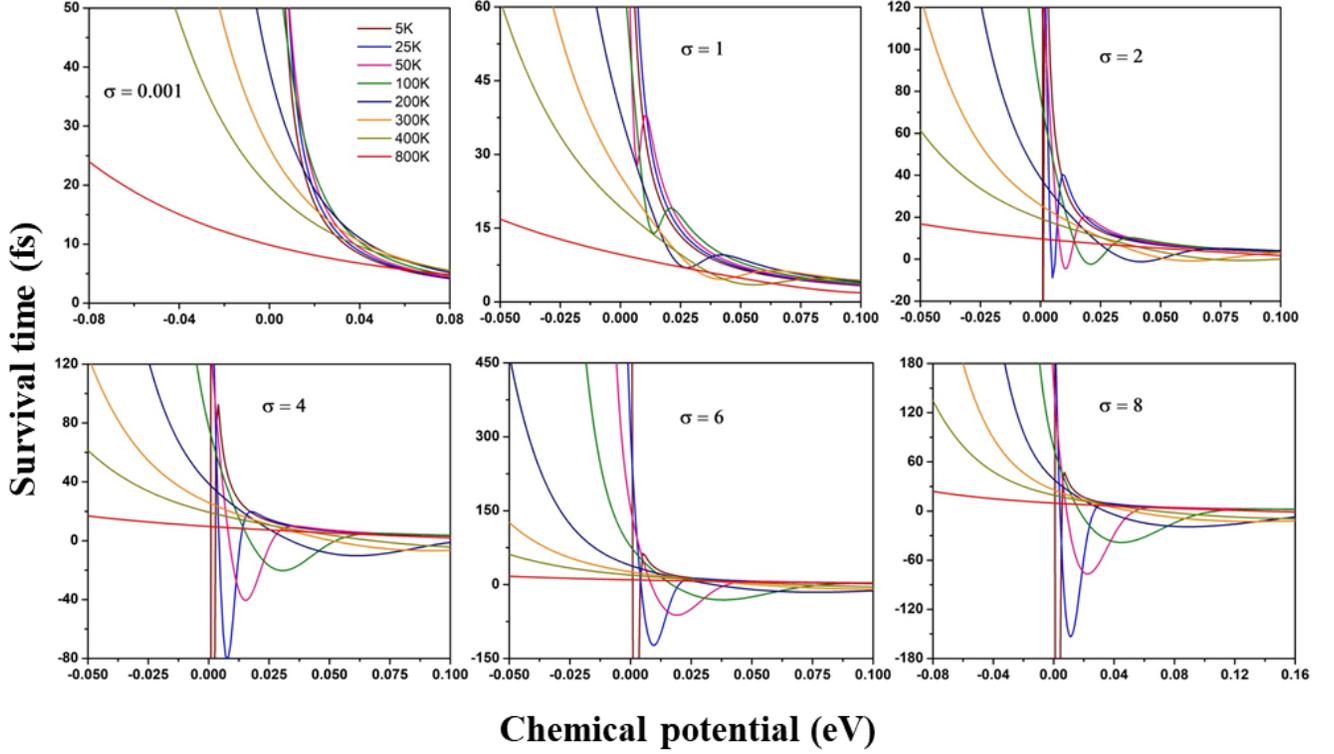

FIG. A8. Entropy weightage on carrier survival time measurement $\left(t = \frac{\hbar \mu_S}{2eD_S}\right)$ generally causes the time delayed D/μ transport. The enhancement of D/μ factor is directly observed while chemical potential increases.

Thus, the electronic compressibility becomes,

$$\frac{\partial n_{2D,S}}{\partial \eta} = \frac{2m}{\pi \hbar^2 \sqrt{2\pi}} \frac{1}{\sigma} \frac{exp\left(\frac{\eta}{k_BT}\right)}{\left(1+exp\left(\frac{\eta}{k_BT}\right)\right)} exp\left(-\frac{\pi^3}{24y}\right) \left\{ \left[1 + exp\left(-\frac{2y^2 exp\left(-\frac{\pi^3}{12y}\right)}{\sigma^2}\right)\right]\left(1 + \frac{\pi^3}{24y}\right) - \left\{y\, exp\left(-\frac{\pi^3}{12y}\right) exp\left(-\frac{2y^2 exp\left(-\frac{\pi^3}{12y}\right)}{\sigma^2}\right)\left[\frac{4y}{\sigma^2} + \frac{\pi^3}{6}\right]\right\}\right\} \quad (D6)$$

where, $y = ln\left(1 + exp\left(\frac{\eta}{k_BT}\right)\right)$.

Inserting Eq. (D5) and (D6) in to Eq. (A21), the D/μ relation for 2D non-equilibrium Schrödinger materials can be derived as,

$$\left(\frac{D}{\mu}\right)_{2D,S} = \frac{k_BT}{e} \left[\frac{y\left(1+exp\left(\frac{\eta}{k_BT}\right)\right)}{exp\left(\frac{\eta}{k_BT}\right)}\right] \left\{ \frac{\left[1+exp\left(-\frac{2y^2 exp\left(-\frac{\pi^3}{12y}\right)}{\sigma^2}\right)\right]}{\left[1+exp\left(-\frac{2y^2 exp\left(-\frac{\pi^3}{12y}\right)}{\sigma^2}\right)\right]\left(1+\frac{\pi^3}{24y}\right) - \left\{y\, exp\left(-\frac{\pi^3}{12y}\right) exp\left(-\frac{2y^2 exp\left(-\frac{\pi^3}{12y}\right)}{\sigma^2}\right)\left[\frac{4y}{\sigma^2} + \frac{\pi^3}{6}\right]\right\}} \right\} \quad (D7)$$

In the case of σ→0 (negligible or zero Gaussian width), the above D/μ equation (Eq. D7) reduced as,

$$\left(\frac{D}{\mu}\right)_{2D,S} = \frac{k_BT}{e} \left[\frac{\left(1+exp\left(\frac{\eta}{k_BT}\right)\right) ln\left(1+exp\left(\frac{\eta}{k_BT}\right)\right)}{exp\left(\frac{\eta}{k_BT}\right)}\right] \left(\frac{1}{1+\frac{\pi^3}{24\, ln\left(1+exp\left(\frac{\eta}{k_BT}\right)\right)}}\right) \quad (D8)$$

*K. Navamani*

In the degenerate cases, $\eta \gg k_B T$, the Eq. (D5), (D6) and (D7) can be revised as,

$$n_{2D,S} = \frac{2m\eta}{\pi\hbar^2\sqrt{2\pi}}\frac{1}{\sigma}exp\left(-\frac{\pi^3 k_B T}{24\eta}\right)\left\{1 + exp\left[-\frac{2\eta^2 exp\left(-\frac{\pi^3 k_B T}{12\eta}\right)}{\sigma^2 k_B^2 T^2}\right]\right\} \quad (D9)$$

$$\frac{\partial n_{2D,S}}{\partial \eta} = \frac{2m}{\pi\hbar^2\sqrt{2\pi}}\frac{1}{\sigma}exp\left(-\frac{\pi^3 k_B T}{24\eta}\right)\left\{\left[1 + exp\left(-\frac{2\eta^2 exp\left(-\frac{\pi^3 k_B T}{12\eta}\right)}{\sigma^2 k_B^2 T^2}\right)\right]\left(1 + \frac{\pi^3 k_B T}{24\eta}\right) - \left\{\frac{\eta}{k_B T}exp\left(-\frac{\pi^3 k_B T}{12\eta}\right)exp\left(-\frac{2\eta^2 exp\left(-\frac{\pi^3 k_B T}{12\eta}\right)}{\sigma^2 k_B^2 T^2}\right)\left[\frac{4\eta}{\sigma^2 k_B T} + \frac{\pi^3}{6}\right]\right\}\right\} \quad (D10)$$

$$\left(\frac{D}{\mu}\right)_{2D,S} = \frac{\eta}{e}\left\{\frac{1 + exp\left(-\frac{2\eta^2 exp\left(-\frac{\pi^3 k_B T}{12\eta}\right)}{\sigma^2 k_B^2 T^2}\right)}{\left[1 + exp\left(-\frac{2\eta^2 exp\left(-\frac{\pi^3 k_B T}{12\eta}\right)}{\sigma^2 k_B^2 T^2}\right)\right]\left(1 + \frac{\pi^3 k_B T}{24\eta}\right) - \left\{\frac{\eta}{k_B T}exp\left(-\frac{\pi^3 k_B T}{12\eta}\right)exp\left(-\frac{2\eta^2 exp\left(-\frac{\pi^3 k_B T}{12\eta}\right)}{\sigma^2 k_B^2 T^2}\right)\left[\frac{4\eta}{\sigma^2 k_B T} + \frac{\pi^3}{6}\right]\right\}}\right\} \quad (D11)$$

For zero Gaussian width (or σ→0), the disorder (or entropy) limited D/μ equation is further reduced as,

$$\left(\frac{D}{\mu}\right)_{2D,S} = \frac{\eta}{e}\left[\frac{1}{1 + \frac{\pi^3 k_B T}{24\eta}}\right] \quad (D12)$$

In pure quantum limit, $T \to 0$,

$$\left(\frac{D}{\mu}\right)_{2D,T\to 0} \equiv \frac{\eta}{e} \cong \frac{E_F}{e} \quad (D13)$$

Now, this relation preserves the earlier D/μ relation (see Eq. A27). In such limit, the diffusion-mobility linearly depends on only the parameter chemical potential. Here, D/μ basically provides one to one correspondence between the electronic information and the transport mechanism of a particular system.

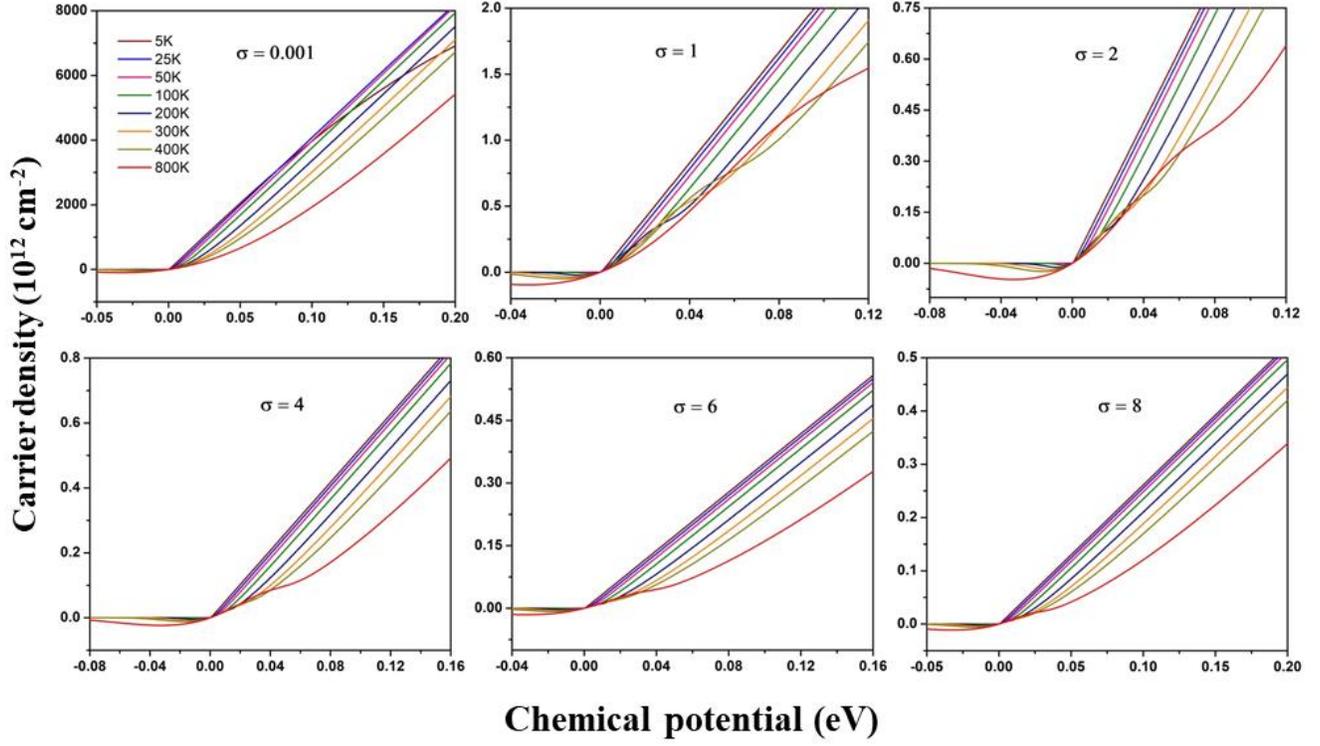

FIG. A9. Entropy modulated carrier density for degenerate materials as a function of chemical potential at different temperature values for different Gaussian variances. The presence of entropy reduces the carrier concentration in the localized region. In the delocalized condition ($\eta > 0$), the measured carrier density enhances with the chemical potential, and decreases with the temperature.

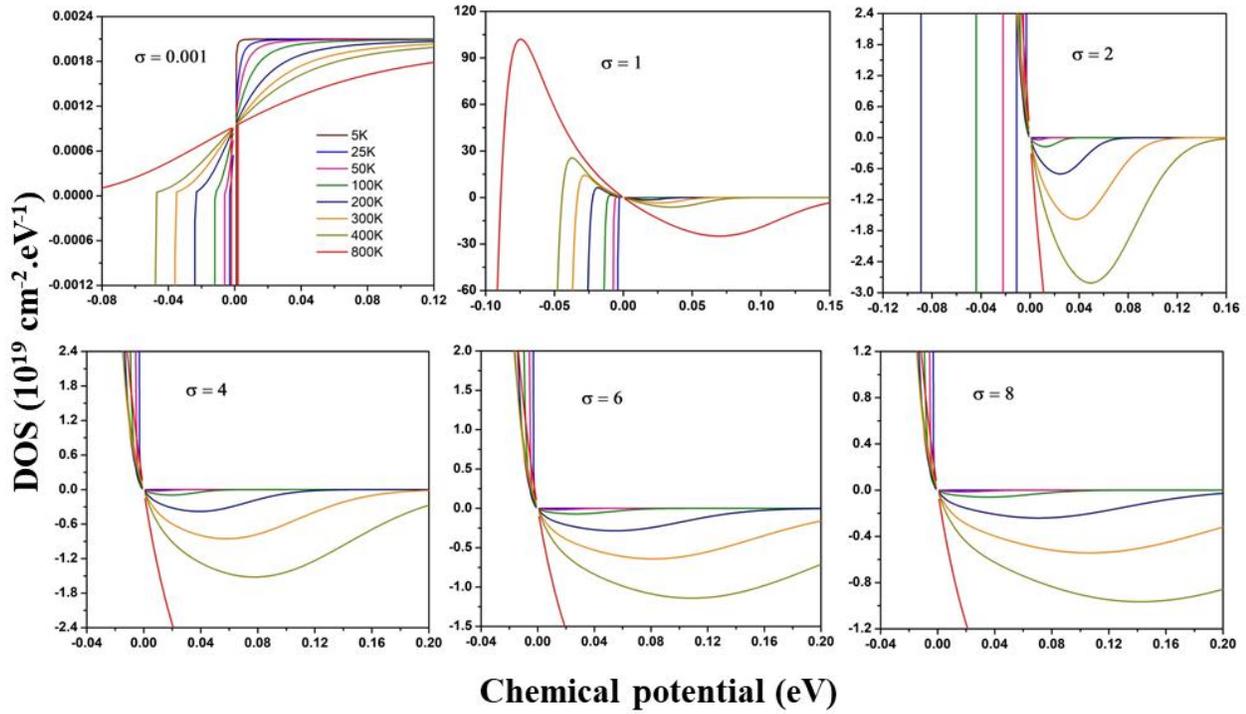

\FIG. A10. Entropy included DOS and its distributions in the wide chemical potential range at different Gaussian variance values. For low Gaussian width, the calculated DOS values are negative in the negative chemical potential values, and the DOS values are positive in the positive chemical potential values. For larger Gaussian width, the above mentioned trend is reversed.

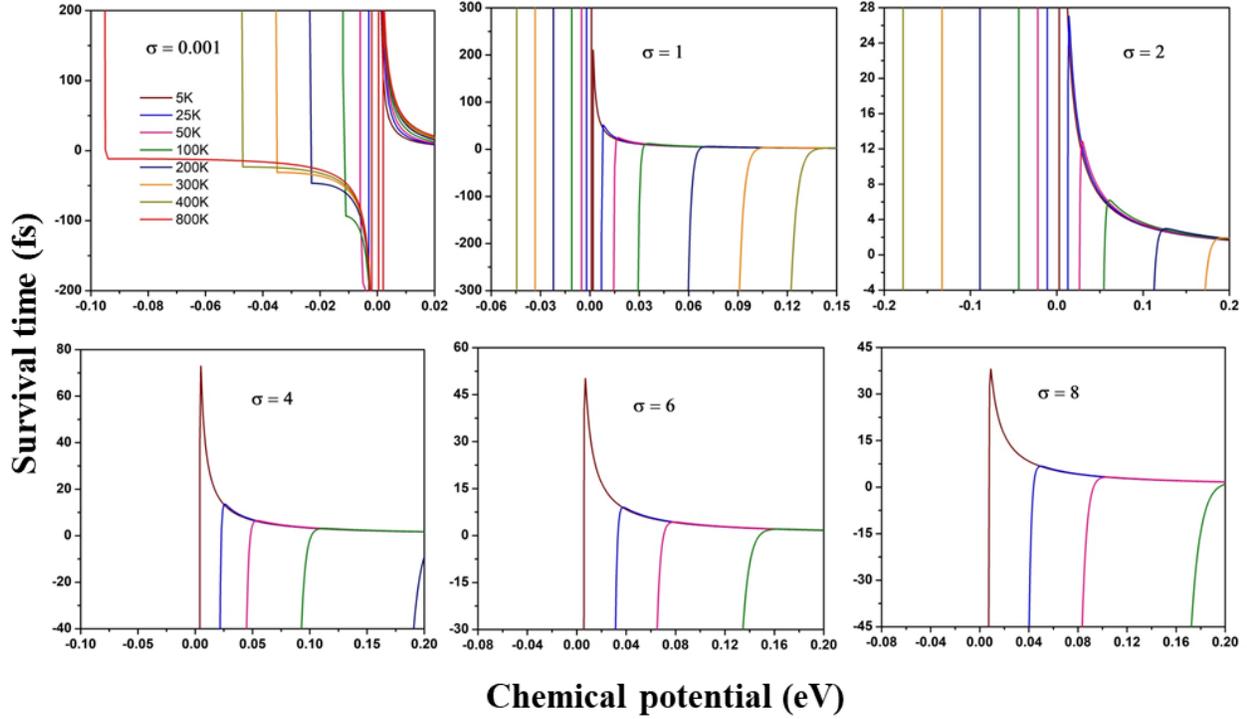

FIG. A11. Entropy effect on carrier survival time plot directly gives the activation chemical potential for D/µ transport in degenerate materials; on the basis of different values of temperature, chemical potential and of Gaussian variance. For instance, at T = 25 K and σ = 8, the expected activation potential is around 40 meV.

## APPENDIX E
**Entropy effect on mobility (via entropy controlled diffusion equation)**

According to Eq. (B14) and (C3), one can be expressed the entropy limited diffusion equation as,

$$D_S = D_0\, exp\left(-\frac{S}{4k_B}\right) = D_0\, exp\left(-\frac{\pi^3}{48\, ln\left(1+exp\left(\frac{\eta}{k_B T}\right)\right)}\right) \tag{E1}$$

In degenerate condition ($\eta \gg k_B T$), the above Eq. (E1) becomes,

$$D_S = D(\eta, T) = D_0\, exp\left(-\frac{\pi^3 k_B T}{48\eta}\right) \tag{E2}$$

To find out the entropy effect on carrier mobility, the Eq. (A23) is divided by Eq. (D8) and now we get the below relation,

$$\frac{(D/\mu)_{S=0}}{(D/\mu)_{S\neq 0}} = \frac{D_0}{D_S}\frac{\mu_S}{\mu_0} = 1 + \frac{\pi^3}{24\, ln\left(1+exp\left(\frac{\eta}{k_B T}\right)\right)} \tag{E3}$$

Substituting the Eq. (E1) in to Eq. (E3), we get

$$\mu_S = \mu(\eta, T) = \mu_0\left[1 + \frac{\pi^3}{24\, ln\left(1+exp\left(\frac{\eta}{k_B T}\right)\right)}\right] exp\left(-\frac{\pi^3}{48\, ln\left(1+exp\left(\frac{\eta}{k_B T}\right)\right)}\right) \tag{E4}$$

For degenerate materials, the above Eq. (E4) is reduced as

$$\mu_S = \mu(\eta, T) = \mu_0\left[1 + \frac{\pi^3 k_B T}{24\eta}\right] exp\left(-\frac{\pi^3 k_B T}{48\eta}\right) \tag{E5}$$

Comparing Eq. (C3) and Eq. (E4), the above entropy dependent mobility Eq. (E4) becomes,

$$\mu_S = \mu_0 \left[1 + \frac{S}{2k_B}\right] exp\left(-\frac{S}{4k_B}\right) \tag{E6}$$

Here, $\mu_0 = \frac{e}{k_B T} \frac{exp\left(\frac{\eta}{k_B T}\right)}{\left(1+exp\left(\frac{\eta}{k_B T}\right)\right) ln\left(1+exp\left(\frac{\eta}{k_B T}\right)\right)} D_0$, and $D_0$ is the diffusion coefficient in the absence of entropy effect, respectively. For degenerate situations, $\mu_0 = \frac{e}{\eta} D_0$.

-----------------------------------------------------------------------------------------------

[1] L. B. Schein and A. R. McGhie, Band-hopping mobility transition in naphthalene and deuterated naphthalene, Phys. Rev. B: Condens. Matter Mater. Phys. **20**, 1631 (1979).

[2] K. Navamani, G. Saranya, P. Kolandaivel, and K. Senthilkumar, Effect of structural fluctuations on charge carrier mobility in thiophene, thiazole and thiazolothiazole based oligomers, Phys. Chem. Chem. Phys. **15**, 17947 (2013).

[3] G. A. H. Wetzelaer, L. J. A. Koster, and P. W. M. Blom, Validity of the Einstein relation in disordered organic semiconductors, Phys. Rev. Lett. **107**, 066605 (2011).

[4] G. A. H. Wetzelaer and P. W. M. Blom, Diffusion-driven currents in organic-semiconductor diodes, NPG Asia Materials **6**, e110 (2014).

[5] Y. Roichman and N. Tessler, Generalized Einstein relation for disordered semiconductors—implications for device performance, Appl. Phys. Lett. **80**, 1948 (2002).

[6] K. Harada, M. Riede, K. Leo, O. R. Hild, and C. M. Elliott, Pentacene homojunctions: Electron and hole transport properties and related photovoltaic responses, Phys. Rev. B **77**, 195212 (2008).

[7] M. G. Ancona, Electron transport in graphene from a diffusion-drift perspective, IEEE Transactions on Electron Devices **57**, 681 (2010).

[8] H.-K. Park and J. Choi, Origin of Voltage-Dependent High Ideality Factors in Graphene-Silicon Diodes, Adv. Electron. Mater. **4**, 1700317 (2018).

[9] S. A. Svatek, et al., Gate tunable photovoltaic effect in MoS$_2$ vertical p-n homostructures, J. Mate. Chem. C **5**, 854 (2017).

[10] C. G. Rodrigues, A. R. Vasconcellos, and R. Luzzi, Nonlinear charge transport in III-N semiconductors: Mobility, diffusion, and a generalized Einstein relation, J. Appl. Phys. **99**, 073701 (2006).

[11] K. Navamani, P. K. Samanta, and S. K. Pati, Theoretical modeling of charge transport in triphenylamine–benzimidazole based organic solids for their application as host-materials in phosphorescent OLEDs, RSC Adv. **8**, 30021 (2018).

[12] K. Navamani, S. K. Pati, and K. Senthilkumar, Site Energy Fluctuation on Charge Transport in Disordered Organic Molecules, (2019).

[13] K. Harada, A. G. Werner, M. Pfeiffer, C. J. Bloom, C. M. Elliott, and K. Leo, Organic homojunction diodes with a high built-in potential: Interpretation of the current-voltage characteristics by a generalized Einstein relation, Phys. Rev. Lett. **94**, 036601 (2005).

[14] B. Kippelen and J.-L. Brédas, Organic photovoltaics, Energy Environ. Sci. **2**, 251 (2009).

[15] Y. An, A. Behnam, E. Pop, G. Bosman, and A. Ural, Forward-bias diode parameters, electronic noise, and photoresponse of graphene/silicon Schottky junctions with an interfacial native oxide layer, J. Appl. Phys. **118**, 114307 (2015).

[16] A. Di Bartolomeo, Graphene Schottky diodes: An experimental review of the rectifying graphene/semiconductor heterojunction, Phys. Rep. **606**, 1 (2016).

[17] D. Sinha and J. U. Lee, Ideal graphene/silicon Schottky junction diodes, Nano lett. **14**, 4660 (2014).

[18] E. A. Henriksen and J. P. Eisenstein, Measurement of the electronic compressibility of bilayer graphene, Phys. Rev. B **82**, 041412 (2010).

[19] E. E. Hroblak, A. Principi, H. Zhao, and G. Vignale, Electrically induced charge-density waves in a two-dimensional electron liquid: Effects of negative electronic compressibility, Phys. Rev. B **96**, 075422 (2017).

[20] K. Zou, X. Hong, and J. Zhu, Effective mass of electrons and holes in bilayer graphene: Electron-hole asymmetry and electron-electron interaction, Phys. Rev. B **84**, 085408 (2011).

[21] J. Martin, N. Akerman, G. Ulbricht, T. Lohmann, J. H. Smet, K. Von Klitzing, and A. Yacoby, Observation of electron–hole puddles in graphene

[22] C. Chen, V. V. Deshpande, M. Koshino, S. Lee, A. Gondarenko, A. H. MacDonald, P. Kim, and J. Hone, Modulation of mechanical resonance by chemical potential oscillation in graphene, Nature Phys. **12**, 240 (2016).

using a scanning single-electron transistor, Nature Phys. **4**, 144 (2008).

[23] B. Skinner, Chemical potential and compressibility of quantum Hall bilayer excitons, Phys. Rev. B **93**, 085436 (2016).

[24] Z. Lin, et al., 2D materials advances: from large scale synthesis and controlled heterostructures to improved characterization techniques, defects and applications, 2D Materials **3**, 042001 (2016).

[25] G. R. Bhimanapati, et al., Recent advances in two-dimensional materials beyond graphene, ACS nano **9**, 11509 (2015).

[26] A. K. Harrison and R. Zwanzig, Transport on a dynamically disordered lattice, Phys. Rev. A **32**, 1072 (1985).

[27] J. Böhlin, M. Linares, and S. Stafstrom, Effect of dynamic disorder on charge transport along a pentacene chain, Phys. Rev. B **83**, 085209 (2011).

[28] S. J. Grimme, Semiempirical GGA-type density functional constructed with a long-range dispersion correction, Comput. Chem. **27**, 1787 (2006).

[29] J. Crossno, et al., Observation of the Dirac fluid and the breakdown of the Wiedemann-Franz law in graphene, Science **351**, 1058 (2016).

[30] J. Chen, X. Y. He, K. H. Wu, Z. Q. Ji, L. Lu, J. R. Shi, J. H. Smet, and Y. Q. Li, Tunable surface conductivity in $Bi_2Se_3$ revealed in diffusive electron transport, Phys. Rev. B **83**, 241304 (2011).

[31] P. T. Brown, et al., Bad metallic transport in a cold atom Fermi-Hubbard system, Science **363**, 379 (2019).

[32] J.-P. Brantut, Transport with strong interactions, Science **363**, 344 (2019).

[33] J. M. Leinaas, Luttinger liquids, Fermi liquids, and fractional statistics, Phys. Rev. B **95**, 155429 (2017).

[34] F. G. Eich, M. Holzmann, and G. Vignale, Effective mass of quasiparticles from thermodynamics, Phys Rev. B **96**, 035132 (2017).

[35] A. L. Fetter and J. D. Walecka, *Quantum theory of many-particle systems* (Dover Publications, INC., 2003).

[36] W. Zhu, V. Perebeinos, M. Freitag, and P. Avouris, Carrier scattering, mobilities, and electrostatic potential in monolayer, bilayer, and trilayer graphene, Phys. Rev. B **80**, 235402 (2009).

[37] N. W. Ashcroft and N. D. Mermin, *Solid State Physics* (Holt, Rinehart and Winston, New York, 1988).

[38] V. S. Khrapai, A. A. Shashkin, M. G. Trokina, V. T. Dolgopolov, V. Pellegrini, F. Beltram, G. Biasiol, and L. Sorba, Filling factor dependence of the fractional quantum Hall effect gap, Phys. Rev. Lett. **100**, 196805 (2008).

[39] A. Troisi, Charge transport in high mobility molecular semiconductors: classical models and new theories, Chem. Soc. Rev **40**, 2347 (2011).

[40] A. Troisi and D. L. Cheung, Transition from dynamic to static disorder in one-dimensional organic semiconductors, J. Chem. Phys. **131**, 014703 (2009).

[41] A. Troisi and G. Orlandi, Charge-transport regime of crystalline organic semiconductors: diffusion limited by thermal off-diagonal electronic disorder, Phys. Rev. Lett. **96**, 086601 (2006).

[42] K. Navamani and K. Senthilkumar, Effect of Structural Fluctuations on Charge Carrier Dynamics in Triazene Based Octupolar Molecules, J. Phys. Chem. C **118**, 27754 (2014).

[43] J. Chen, et al., Gate-voltage control of chemical potential and weak antilocalization in $Bi_2Se_3$, Phys. Rev. Lett. **105**, 176602 (2010).

[44] K. Lee, B. Fallahazad, J. Xue, D. C. Dillen, K. Kim, T. Taniguchi, K. Watanabe, and E. Tutuc, Chemical potential and quantum Hall ferromagnetism in bilayer graphene, Science **345**, 58 (2014).

[45] L. Li, N. Lu, M. Liu, and H. Bässler, General Einstein relation model in disordered organic semiconductors under quasiequilibrium, Phys. Rev. B **90**, 214107 (2014).

[46] D. L. Cheung and A. Troisi, Modelling charge transport in organic semiconductors: from quantum dynamics to soft matter., Phys. Chem. Chem. Phys. **10**, 5941 (2008).

[47] L. Fallani and M. Inguscio, Controlling Cold-Atom Conductivity, Science **322**, 1480 (2008).

[48] S. Ciuchi and S. Fratini, Electronic transport and quantum localization effects in organic semiconductors, Phys. Rev. B **86**, 245201 (2012).

[49] E. M. Hajaj, O. Shtempluk, V. Kochetkov, A. Razin, and Y. E. Yaish, Chemical potential of inhomogeneous single-layer graphene, Phys. Rev. B **88**, 045128 (2013).

[50] H. Yamamoto, H. Kasajima, W. Yokoyama, H. Sasabe, and C. Adachi, Extremely-high-density carrier injection and transport over 12000 A/$cm^2$ into organic thin films, Appl. Phys. Lett. **86**, 083502 (2005).

[51] D. J. Griffiths and D. F. Schroeter, *Introduction to quantum mechanics* (Cambridge University Press, 2018).

[52] Y. A. Berlin, F. C. Grozema, L. D. A. Siebbeles, and M. A. Ratner, Charge transfer in donor-bridge-acceptor systems: Static disorder, dynamic

fluctuations, and complex kinetics, J. Phys. Chem. C **112**, 10988 (2008).

[53] K. Navamani and K. Senthilkumar, Effect of Dynamic Disorder on Charge Transport in Organic Molecules, cond-mat.soft **1705.05648** (2017).